\documentclass[12pt,onecolumn,journal]{IEEEtran}
\usepackage{cite}
\usepackage{setspace}
\usepackage{multicol}

\usepackage[utf8x]{inputenc}
\usepackage[bookmarks=false]{hyperref}
\usepackage{graphicx}
\usepackage{times}
\usepackage{xpatch}
\usepackage{textcomp}
\usepackage{float}
\usepackage{url}

\usepackage{amssymb}
\usepackage{amsthm}
\usepackage{amsxtra}
\usepackage{amsmath}
\usepackage{algorithmic}
\usepackage{cite}
\usepackage{array}
\usepackage{bigints}
\usepackage{mathtools}
\usepackage{enumerate}
\usepackage{verbatim}
\usepackage{epsfig}
\usepackage{caption}
\usepackage{subcaption}
\usepackage{breqn}
\usepackage{color}
\usepackage{yfonts}
\usepackage{atbegshi}
\usepackage{mathrsfs}
\usepackage{soul}

\allowdisplaybreaks

\DeclareMathOperator*{\argmin}{arg\,min}

\newtheorem{theorem}{Theorem}

\newtheorem{proposition}[theorem]{Proposition}

\newtheorem*{remark*}{Remark}

\newcommand{\blue}{\textcolor{black}}

\title{Reconfigurable Adaptive Channel Sensing}
\author{
\IEEEauthorblockN{Manuj Mukherjee$\dag$}\hspace {1cm} \and \IEEEauthorblockN{Aslan Tchamkerten$^*$} \hspace {1cm} \and \IEEEauthorblockN{Chadi Jabbour$^*$}  
}

\begin{document}
%\history{}
%\doi{}

\maketitle
%\author{\uppercase{Manuj Mukherjee}\authorrefmark{1}, \IEEEmembership{Member, IEEE},
%\uppercase{Aslan Tchamkerten}\authorrefmark{1}, \IEEEmembership{}, and \uppercase{Chadi Jabbour}\authorrefmark{1}, \IEEEmembership{}}
%\address[1]{D\'{e}partement COMELEC, Telecom ParisTech, Paris, France.}
%\tfootnote{}

%\markboth
%{Mukherjee \headeretal: AdaSense: An Adaptive Low Energy Channel Sensing Scheme}
%{Mukherjee \headeretal: AdaSense: An Adaptive Low Energy Channel Sensing Scheme}

%\corresp{Corresponding author: Manuj Mukherjee (e-mail: manuj.mukherjee@telecom-paristech.fr).}

\renewcommand{\thefootnote}{}
\footnotetext{
\noindent $^*$ A. Tchamkerten and C. Jabbour are with Telecom Paris, Institut Polytechnique de Paris, France. Email: \{aslan.tchamkerten, chadi.jabbour\}@telecom-paris.fr.\\
\noindent $\dag$ M. Mukherjee is with Bar Ilan University, Ramat Gan, Israel. Email: mukherm@biu.ac.il}

\renewcommand{\thefootnote}{\arabic{footnote}}

\begin{abstract}
      Channel sensing consists of probing the channel from time to time to check whether or not it is active—say, because of an incoming message. When communication is sparse with  information being sent once in a long while, channel sensing becomes a significant source of energy consumption. How to reliably detect messages while minimizing the receiver energy consumption? This paper addresses this problem through a reconfigurable scheme, referred to as AdaSense, which exploits the dependency between the receiver noise figure (\emph{i.e.}, the receiver added noise) and the receiver power consumption; a higher power typically translates into less noisy channel observations. AdaSense begins in a low power low reliability mode and makes a first tentative decision based on a few channel observations. If a message is declared, it switches to a high power high reliability mode to confirm the decision, else it sleeps for the entire duration of the second phase. Compared to prominent detection schemes such as the BMAC protocol, AdaSense provides relative energy gains that grow unbounded in the small probability of false-alarm regime, as communication gets sparser. In the non-asymptotic regime energy gains are 30\% to 75\% for communication scenarios typically found in the context of wake-up receivers.
\iffalse
      Channel sensing is a fundamental task of modern radio receivers and consists of probing the channel from time to time to check whether or not it is active---say, because of an incoming message. \iffalse Channel sensing should be efficient in terms of error probabilities, false-alarm and miss-detection, and energy consumption. \fi
            When communication is sparse with small amounts of information sent once in a long while, channel sensing becomes a relatively significant source of energy consumption. How to reliably detect messages while minimizing the receiver energy consumption? This paper addresses this problem through a simple reconfigurable scheme, referred to as AdaSense, which exploits the dependency between the receiver noise figure (that is the receiver added noise) and the receiver power consumption; a higher power typically translates into less noisy channel observations. AdaSense begins in a low power low reliability mode and makes a first tentative decision based on a few channel observations. If a message is declared, it switches to a high power high reliability mode to confirm the decision,  else it sleeps for the entire duration of the second phase. When compared to prominent detection schemes such as the BMAC protocol, AdaSense provides relative energy gains that grow unbounded in the small probability of false-alarm regime, as communication gets sparser. In the non-asymptotic regime energy gains are $30\%$ to $75\%$ for a variety of practical communication scenarios typically found in the context of wake-up receivers. \fi
\end{abstract}

%\begin{keywords}
%Channel sensing, energy efficient communication, low-power communication, wake-up receiver, wireless sensor networks.
%\end{keywords}

%\titlepgskip=-15pt


\section{Introduction}\label{sec:intro}

To minimize energy consumption modern receivers are typically \emph{duty cycled} \cite{HC02, El02, WMAC, SMAC, TMAC, BMAC, XMAC}, {\it{i.e.}}, they listen only at predetermined time periods\iffalse---the time interval between the listening periods should be set so that to achieve a target delay \cite[Section~I.A]{survey}\fi. \iffalse Current wake-up receivers are typically duty-cycled \cite{WUDC, WUDC2, WUDC3, WUDC4}. \fi
\iffalse If no message is detected, the receiver goes back to sleep, and otherwise \blue{depending on the application, it either takes the necessary action (like opening the smart lock), or stays awake to receive the forthcoming message (like receiving the location of the forest fire). However, the time spent in listening cannot be reduced arbitrarily, since the listening periods must be adequately long to reliably detect a preamble. Alternately, the sleep periods cannot be lengthened arbitrarily without increasing the delay in detecting the preamble. One can therefore optimize the sleep/listening schedules to minimize energy consumption given a tolerable delay \cite{7890362,7870667}.} \fi 
To alert the receiver of an incoming message, the transmitter sends a preamble either at the beginning of a listening period ({\it{e.g.}}, \cite{HC02, El02, WMAC, WUDC4}), or immediately in which case the preamble is long enough to cover at least one listening period ({\it{e.g.}}, \cite{BMAC, WUDC2}).
\iffalse During the listening phase the receiver performs a hypothesis test based on a fixed number of samples and decides whether or not a preamble is present. We refer to such schemes as ``single-phase.'' In this paper we propose a new adaptive channel sensing scheme, referred to as AdaSense (for adaptive sensing), which outperforms single-phase  schemes in terms of energy consumption given target error probabilities.\fi
In both cases the receiver performs a binary hypothesis test during the listening periods to decide whether or not a preamble is present. 

How to minimize the energy consumption of a sensing scheme for a given  reliability level specified in terms of the probabilities of false-alarm and miss-detection? In this paper we address this question through a simple sensing scheme referred to as AdaSense (for adaptive sensing). In a nutshell, AdaSense reconfigures its noise level in an adaptive manner so that to strike a balance between reliability and energy consumption.
In more details, a significant portion of the noise in a machine-to-machine communication channel is due to the \emph{receiver noise},\footnote{{See Section~\ref{sec:compfin} for details.}} which is typically a non-increasing function of the power consumption at the receiver (see, {\it{e.g.}}, \cite[Chapter~12]{Lee});  higher reliability requires greater power consumption. AdaSense exploits this dependency to adaptively choose the noise level of the observed samples so that to minimize the overall energy consumed in channel sensing.

\iffalse

Second, the receiver's decision need not be made after observing all $n$ samples of the listening period but can be made after observing only say $N\leq n$ samples, where $N$ is random and chosen based on the confidence level reached so far. 
\fi

\iffalse
\begin{figure}
\centering
\resizebox{0.2\textwidth}{!}{\includegraphics{DC1.eps}}
\caption{Duty cycling}
\label{fig:dc}
\end{figure}
\fi 

AdaSense has two phases. In the first phase, AdaSense observes a small batch of samples at low power and makes a tentative decision on whether or not a preamble is present. If no preamble is detected AdaSense  stops, declares that there is no preamble, and moves to the next listening phase. If a preamble is declared, AdaSense enters a second confirmation phase, and observes a fresh batch of samples at a higher power. At the end of this second phase, AdaSense decides whether or not the preamble is present. AdaSense was inspired by the adaptive detection schemes proposed in \cite{banerjee2013data,tchamkerten2014energy,ebrahimzadeh2015sequential,AV18} which aim at minimizing the number of samples needed to efficiently detect a message. While the number of samples can be taken as a proxy for energy consumption, the present paper goes further by considering the actual energy spent in the observation of a sample through its dependency with the receiver noise figure. In particular, leveraging upon this dependency AdaSense  adaptively varies the power consumption, in addition to the number of samples.
\iffalse
To summarize, an energy efficient channel sensing scheme should ideally have the following characteristics:
\begin{itemize}
\item Ability to make decision with a fewer number of samples.
\item Work on batches of samples, instead of individual ones. 
\item Adapt the power consumption in between batches of samples, to suitably exploit the noise power versus power consumption tradeoff.
\end{itemize}
In this work, we design a new two-phase channel sensing scheme referred to as AdaSense which operates as follows.  satisfying all the three properties listed above.
\fi

We compared AdaSense against two well-known schemes, the ``single-phase'' scheme and the \emph{clear channel assessment algorithm} of the \emph{Berkeley Media Access Control} (BMAC) protocol, henceforth referred to as the BMAC scheme \cite{BMAC}. The single-phase scheme refers to the standard binary hypothesis test where an optimal decision is taken based on the observations at constant power of a fixed number of $n$ samples, where $n$ denotes the length of the preamble. If a preamble is declared, the receiver stops and otherwise it waits for the next listening period. The BMAC scheme instead is sequential. The receiver observes samples at a constant power and declares that the preamble is present only if all $n$ sample values exceed a given threshold. As soon as one sample value is below the threshold, it stops, declares that no preamble is present, and waits for the next listening period.

The rest of the paper is organized as follows. In Section~\ref{sec:model}, we give a precise description of the channel sensing problem. In Section~\ref{sec:ada}, we describe AdaSense and compare it with the single-phase scheme and the BMAC scheme. We first make analytical comparisons in the small probability of false-alarm regime when communication gets sparser, then make comparisons in non-asymptotic regimes through numerical performance evaluations. In Section ~\ref{sec:imp}, we discuss a practical implementation of AdaSense. Finally, in Section~\ref{sec:conc} we draw a few concluding remarks.

\iffalse
------------

only after observing all of these samples. We refer to such schemes as ``single-phase.'' Alternately, the receivers can adaptively choose to stop and make a decision after observing only a few of these samples, {\it{e.g.}}, in the

Second, the BMAC scheme makes its decision based on individual samples. Hence, a lone extremely noisy sample will have a detrimental effect on the decision. To mitigate such effects the noise level needs to be reduced which impacts power consumption. Alternatively, decisions need to be made on batches of samples instead of individual samples.

------------
\fi

\section{Channel sensing: reliability and energy consumption}\label{sec:model}

\iffalse In a variety of wireless sensor networks communication is sparse with messages transmitted once in a long while. To help the receiver detect information transmission, a message is preceded by a preamble. We consider a receiver that senses the channel at predetermined time instances and estimates whether a preamble is present or not. Depending on the protocol, the preamble can either be sent at one of the receiver's wake up instants ({\it{e.g.}}, \cite{HC02, El02, WMAC}, \cite{WUDC3}), or is sent at a random time but is long enough to ensure that the receiver can observe sufficiently many samples from the preamble ({\it{e.g.}}, in \cite{BMAC}, \cite[Section~2]{WUDC2}). In both cases there is a maximum number of $n\geq 1$ samples that the receiver can observe per wake-up before deciding on whether or not a preamble is present. The receiver can choose to observe these $n$ samples to maximize reliability, or can observe fewer samples and compromise between reliability and energy consumption. \fi

\iffalse
\begin{figure}
\centering
\resizebox{0.3\textwidth}{!}{\includegraphics{channel.eps}}
\caption{Channel model}
\label{fig:channel}
\end{figure}
\fi

Channel sensing at the physical layer amounts to a binary hypothesis test. Consider $N$ samples $Y_1,Y_2,\ldots,$ $Y_N$ that represent the outputs of a coherent receiver that observes a modulated binary message $M\in \{0,1\}$, repeated $N$ times, and corrupted by additive noise. \iffalse \blue{as well as slow fading}.  \footnote{{We restrict ourselves to binary modulation for simplicity of exposition and since it is widely used in practice, as in the context of wake-up receivers (see, {\it{e.g.}}, \cite{survey}). Our analysis easily extends to higher order modulations and is omitted.}} \fi  Hence,  
\begin{equation}
Y_i=M\cdot\sqrt{P}+Z_i, \: \: 1\leq i\leq N,\label{eq:channel}
\end{equation}
where $Z_i\sim \mathcal{N}(0,\sigma^2_i)$ denotes the noise of the $i$-th sample, and where $P$ denotes the received power.\iffalse \footnote{We consider a slow fading channel model. \blue{The received power $P$ is given by $P=|h|^2P_t$, where $P_t$ is the transmit power and the fading coefficient is denoted by $h$. Channel state information at the receiver is assumed, and hence $P$ is known to the receiver.}} \fi 

The two possible values of $M$ are interpreted as the hypothesis $H_0=\{M=0\}$, corresponding to no preamble, and $H_1=\{M=1\}$, corresponding to a preamble being present. These hypothesis are supposed to have a known prior which reflects the level of communication sparsity $$p_1=\text{Pr}(H_1)=1-\text{Pr}(H_0).$$ 
\iffalse, where
\[
X=
\begin{cases}
\sqrt{P} & \text{ if }M=1\\
0 & \text{ if }M=0,
\end{cases}
\]
and where $P$ denotes the received power.  

Hence, the $Y_i$'s are i.i.d. and distributed according to
\begin{equation}
Y_i\sim
\begin{cases}
\mathcal{N}(0,\sigma^2) & \text{ if }X=0\\
\mathcal{N}(\sqrt{P},\sigma^2) & \text{ if }X=\sqrt{P}.
\end{cases}
\label{eq:sdist}
\end{equation} 
\fi 

Based on $Y_1,Y_2,\ldots,Y_N$ the receiver provides a message estimate $\widehat{M}\in \{0,1\}$. 
\iffalse We remark here that while our channel model is discrete time, it is indeed a valid model for continuous signals. This is because, in coherent receivers, the signal after carrier removal is passed through a rectifier, a low noise amplifier, and an integrator. The output of these operations are the samples $Y_i$s given by \eqref{eq:channel}. Then, based on these $Y_i$s, the receiver makes a decision between $H_0$ and $H_1$. \fi
The reliability of the estimator is quantified in terms of the probabilities of false-alarm and  miss-detection
\begin{align}
P_{\text{FA}} & =\text{Pr}(\widehat{M}=1|H_0)\notag\\
P_{\text{Miss}} & =\text{Pr}(\widehat{M}=0|H_1).\label{eq:rel}
\end{align}
In addition to reliability, we are interested in the average energy $E$ spent by the receiver in observing $Y_1,\ldots, Y_N$. \blue{The parameters of the receiver circuit that have a significant impact on the energy consumption, include the central frequency, the receiver architecture, the noise figure, \textit{etc.} Nevertheless, for a given application with a fixed central frequency, received power, and data rate, the receiver energy consumption is primarily determined by the receiver noise figure.} This is  quantified as follows. First note that the noise variance $\sigma^2_i$ may be decomposed as $$\sigma^2_i=\sigma^2_t+\sigma_{r,i}^2$$ where $\sigma^2_t$ denotes the contribution due to thermal noise and where $\sigma_{r,i}^2$ denotes the contribution of the receiver. In many practical setups (see Section~\ref{sec:compfin}), the contribution of the thermal noise $\sigma_t^2$ is negligible with respect to the receiver noise and is ignored. Therefore, we assume that $$\sigma_i^2=\sigma_{r,i}^2.$$

The receiver noise variance $\sigma_{r,i}^2$ is typically a function of the receiver power consumption \blue{$P_{r,i}$}; the lower $\sigma_{r,i}^2$ the larger $P_{r,i}$. This function depends on the receiver circuit itself, and particularly on the low noise amplifier used in the receiver circuit (see, {\it{e.g.}}, \cite[Chapter~12]{Lee}). We model this dependency by letting $$\sigma_{r,i}^2=f(P_{r,i})$$ where $f(\cdot)$ denotes a non-negative and non-increasing function, which we shall refer to as the \emph{noise profile} of the receiver---in practice this function can be determined by means of electrical simulations on the low noise amplifier.\footnote{We emphasize the distinction between $P$, the power of the received signal, and $P_r$ the power consumed by the receiver to observe the signal.} Throughout the paper we assume that $$\sigma_{r,i}^2=k P_{r,i}^{-\gamma}$$ for some known $k> 0$ and $\gamma\geq 1 $---the case $\gamma<1$ is arguably less natural and shall be omitted.

Our results (next section) immediately extend to the case where $\sigma^2_t$ is no longer negligible, but yield slightly more cumbersome results---this is briefly alluded to in Section~\ref{sec:conc}.

Without loss of generality, the number of observed samples $N$ is at most equal to $n$, the length of the preamble. Moreover, $N$ is allowed to causally depend on past observed samples. For instance, after observing a few samples $Y_i$ that clearly indicate the presence of the message, the receiver may stop and output $\widehat{M}=1$.  The power $P_{r,i}$ at which symbol $Y_i$ is observed may depend on past observed samples $Y_1,Y_2,\ldots,Y_{i-1}$.\footnote{In probability language, $N$ is a stopping time defined on the natural filtration induced by the process $Y_1,Y_2,\ldots$ and process $P_{r,1}, P_{r,2},\ldots$ is predictable with respect to this filtration.}  The average energy consumption till a decision is made is then given by
\begin{equation}
    E={\mathbb{E}}\left(\sum_{i=1}^N P_{r,i}\right)\label{eq:ave}
\end{equation}
where ${\mathbb{E}}$ denotes expectation over the channel noise and over the two hypothesis $H_1$ and $H_0$, assumed to have prior  $p_1$ and $1-p_1$, respectively. Here we assume that each symbol has unit duration, without loss of generality.\footnote{For an arbitrary symbol duration $T_s$, just multiply $E$ by $T_s$ in equation \eqref{eq:ave}.}

 \section{\blue{Results}}\label{sec:ada}
We first describe \emph{AdaSense} in Section~\ref{descr}, a scheme which aims at minimizing $E$ for fixed $P_{\text{FA}}$ and~$P_{\text{Miss}}$. We then compare AdaSense against two well-known detection schemes, namely the single-phase scheme and the BMAC scheme. In Section~\ref{asym}, we provide analytical comparisons in the regime of small probability of false-alarm and sparse communication, which is relevant for IoT type of applications. In Section~\ref{sec:compfin}, we numericaly compare these schemes for a variety of non-asymptotic parameters.   

\subsection{\blue{Description of AdaSense}}\label{descr}
\begin{figure}
\centering
\resizebox{0.4\textwidth}{!}{\includegraphics{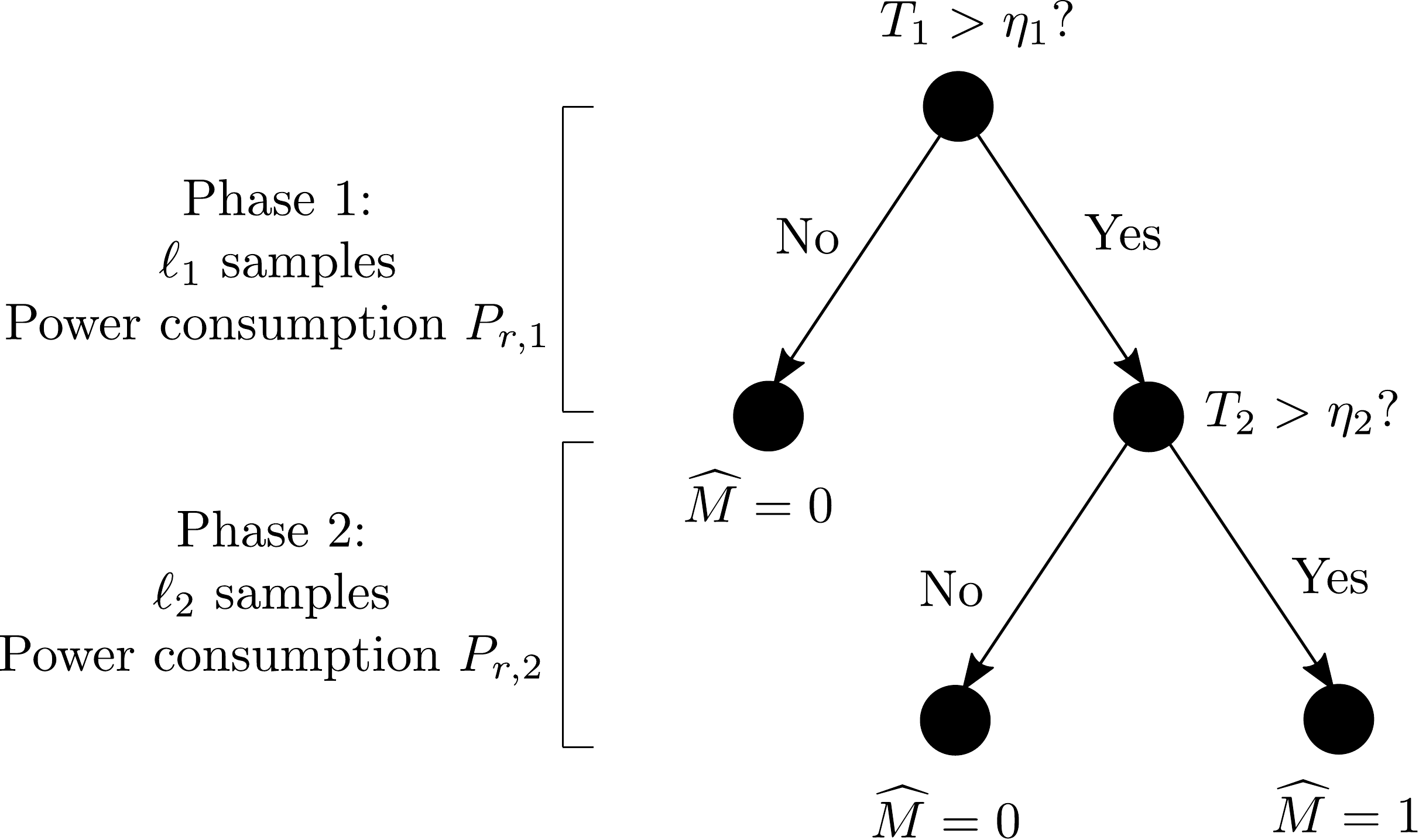}}
\caption{AdaSense}
\label{fig:2p}
\end{figure}

 AdaSense is a two-phase scheme as illustrated in Fig.~\ref{fig:2p}. It starts by observing a first batch of samples and makes a tentative decision. If $H_0$ is declared AdaSense stops, and if $H_1$ is declared AdaSense performs a second ``highly reliable'' confirmation test based on a second batch of samples. If hypothesis $H_1$ is confirmed, AdaSense outputs $H_1$, else it outputs $H_0$. Details follow.   
 
 Let $\ell_1$ and $\ell_2$ be two nonnegative integers such that $\ell_1+\ell_2\leq n$, and let $P_{r,1}, P_{r,1}$ be nonnegative constants. The receiver starts by observing $\ell_1$ samples $Y_1,Y_2,\cdots,Y_{\ell_1}$ at a constant power per sample $P_{r,1}$. It then performs a standard log-likelihood ratio (LLR) test to discriminate between the hypothesis $H_0$ and $H_1$ (see Case~III.B.2 of \cite{HVP}), {\it{i.e.}}, the receiver checks whether 
$$
T_1\triangleq\frac{\sqrt{P}}{\sigma_{r,1}^2}\sum_{i=1}^{\ell_1}Y_i-\frac{\ell_1P}{2\sigma_{r,1}^2}\leq \eta_1,
$$
for some given threshold $\eta_1$, and where $\sigma_{r,1}^2=f(P_{r,1})$ . If this inequality is satisfied, the receiver stops and declares $\widehat{M}=0$. Otherwise, it enters a second confirmation phase, observes the next $\ell_2$ samples {$Y_{\ell_1+1},\ldots, Y_{\ell_1+\ell_2}$} at a constant power per sample $P_{r,2}$, typically greater than $P_{r,1}$, and performs a second LLR test on this second batch of samples, {{\it{i.e.}}}, it checks whether 
$$
T_2\triangleq\frac{\sqrt{P}}{\sigma_{r,2}^2}\sum_{i=\ell_1+1}^{\ell_1+\ell_2}Y_i-\frac{\ell_2P}{2\sigma_{r,2}^2}>\eta_2
$$
or some given threshold $\eta_2$, and where $\sigma_{r,2}^2=f(P_{r,2})$. If this second inequality is satisfied, the receiver declares $\widehat{M}=1$, and otherwise $\widehat{M}=0$. 

For this scheme, $P_{\text{FA}}$, $P_{\text{Miss}}$ and $E$ are explicitly given by (see Appendix~\ref{app:exact})\footnote{$Q$ refers to the standard $Q$-function defined as $Q(x)=\frac{1}{\sqrt{2\pi}}\int_{x}^\infty e^{-\frac{u^2}{2}}du$.} 

\begin{equation}
P_{\text{FA}}=Q\biggl(\frac{\sigma_{r,1}\eta_1}{\sqrt{\ell_1P}}+\frac{1}{2}\sqrt{\frac{\ell_1P}{\sigma_{r,1}^2}}\biggr)Q\biggl(\frac{\sigma_{r,2}\eta_2}{\sqrt{\ell_2P}}+\frac{1}{2}\sqrt{\frac{\ell_2P}{\sigma_{r,2}^2}}\biggr), \label{eq:Ada:pfa}    
\end{equation}

\begin{equation} 
P_{\text{Miss}} = Q\biggl(-\frac{\sigma_{r,1}\eta_1}{\sqrt{\ell_1P}}+\frac{1}{2}\sqrt{\frac{\ell_1P}{\sigma_{r,1}^2}}\biggr)+\left(1-Q\biggl(-\frac{\sigma_{r,1}\eta_1}{\sqrt{\ell_1P}}+\frac{1}{2}\sqrt{\frac{\ell_1P}{\sigma_{r,1}^2}}\biggr)\right) Q\biggl(-\frac{\sigma_{r,2}\eta_2}{\sqrt{\ell_2P}}+\frac{1}{2}\sqrt{\frac{\ell_2P}{\sigma_{r,2}^2}}\biggr),\label{eq:Ada:pmiss}
\end{equation}
and 
\begin{align}
E=\ell_1P_{r,1}+p_c\ell_2P_{r,2}, \label{eq:Ada:en}
\end{align}
where $p_c$ denotes the probability of having a second phase and is given by
\begin{equation*}
{p_c} \overset{{def}}{=} (1-p_1) Q\biggl(\frac{\sigma_{r,1}\eta_1}{\sqrt{\ell_1P}}+\frac{1}{2}\sqrt{\frac{\ell_1P}{\sigma_{r,1}^2}}\biggr)+p_1\biggl(1-Q\biggl(-\frac{\sigma_{r,1}\eta_1}{\sqrt{\ell_1P}}+\frac{1}{2}\sqrt{\frac{\ell_1P}{\sigma_{r,1}^2}}\biggr)\biggr).
\end{equation*}

Given ${P}_{\text{FA}}\leq \alpha$, ${P}_{\text{Miss}}\leq \beta$, the received power $P$, the preamble length $n$ and the sparsity level $p_1$, the set of parameters $\{\ell_1,P_{r,1},\eta_1,\ell_2,P_{r,2},\eta_2\}$ are chosen so that to minimize the average energy:
\begin{equation}
\{\ell_1^*,P_{r,1}^*,\eta_1^*,\ell_2^*,P_{r,2}^*,\eta_2^*\}=\displaystyle\argmin_{{\substack{\{\ell_1,P_{r,1},\eta_1,\ell_2,P_{r,2},\eta_2\}:\\\ell_1+\ell_2\leq n\\P_{\text{FA}}\leq \alpha\\{P}_{\text{Miss}}\leq \beta}}}E.\label{eq:opt}
\end{equation}

The above optimization problem admits no analytical solution but can easily be numerically evaluated as described in Section~\ref{sec:compfin}. It should also be noted that this optimization needs to be carried only once per target values $\alpha$, $\beta$, $n$, $p_1$ and $P$. Depending on $(n,\alpha,\beta)$ there might be no solution\footnote{That is no $\{\ell_1,P_{r,1},\eta_1,\ell_2,P_{r,2},\eta_2\}$ such that $\ell_1+\ell_2\leq n$, $P_{\text{FA}}\leq \alpha$, and ${P}_{\text{Miss}}\leq \beta$} in which case we set $E=\infty$. However, if only $(\alpha,\beta)$ are fixed and $n$ can be optimized over then the above optimization problem always admits a feasible solution {(see, \textit{e.g.}, \cite[Example II.D.1]{HVP})}.

\subsection{\blue{Performance: Small probability of False-Alarm and Sparse Communication regime}}\label{asym}
The next result explicits a tradeoff between $E$ and $P_{\text{FA}}$ at fixed  $P_{\text{Miss}}$ for AdaSense, when $P_{\text{FA}}$ tends to zero (the proof is deferred to Appendix~\ref{app:asympada}):\footnote{We make use of the Big O notation for limiting function behavior.}
\begin{theorem}[AdaSense, asymptotics]\label{th:asympada}
Suppose $f(P_r)=kP_r^{-\gamma}$ for some fixed $\gamma\geq 1$ and $k>0$. For any fixed  $p_1,P>0$, $0<\beta\leq 1$ and $0<\varepsilon<1$ AdaSense achieves $P_{\text{Miss}}\leq\beta$ and 

\begin{align}
 E&\leq \frac{2k p_1  }{P(1-\varepsilon-o(1))}\ln\biggl(\frac{1}{P_{\text{FA}}}\biggr)
 \end{align}
where $o(1)\to 0$ as $P_{\text{FA}}\to 0$. Moreover, this tradeoff between $P_{\text{Miss}}$, $P_{\text{FA}}$, and $E$ is achievable with $n=O(\ln(1/P_{\text{FA}}))$.
\end{theorem}
The important thing to notice here is that the sparser the communication, that is the smaller $p_1$, the lower $E$. To put this result into perspective we now consider two other widely known detection schemes, namely the basic ``single-phase'' scheme and the BMAC scheme.

\begin{figure}
\centering
\resizebox{0.4\textwidth}{!}{\includegraphics{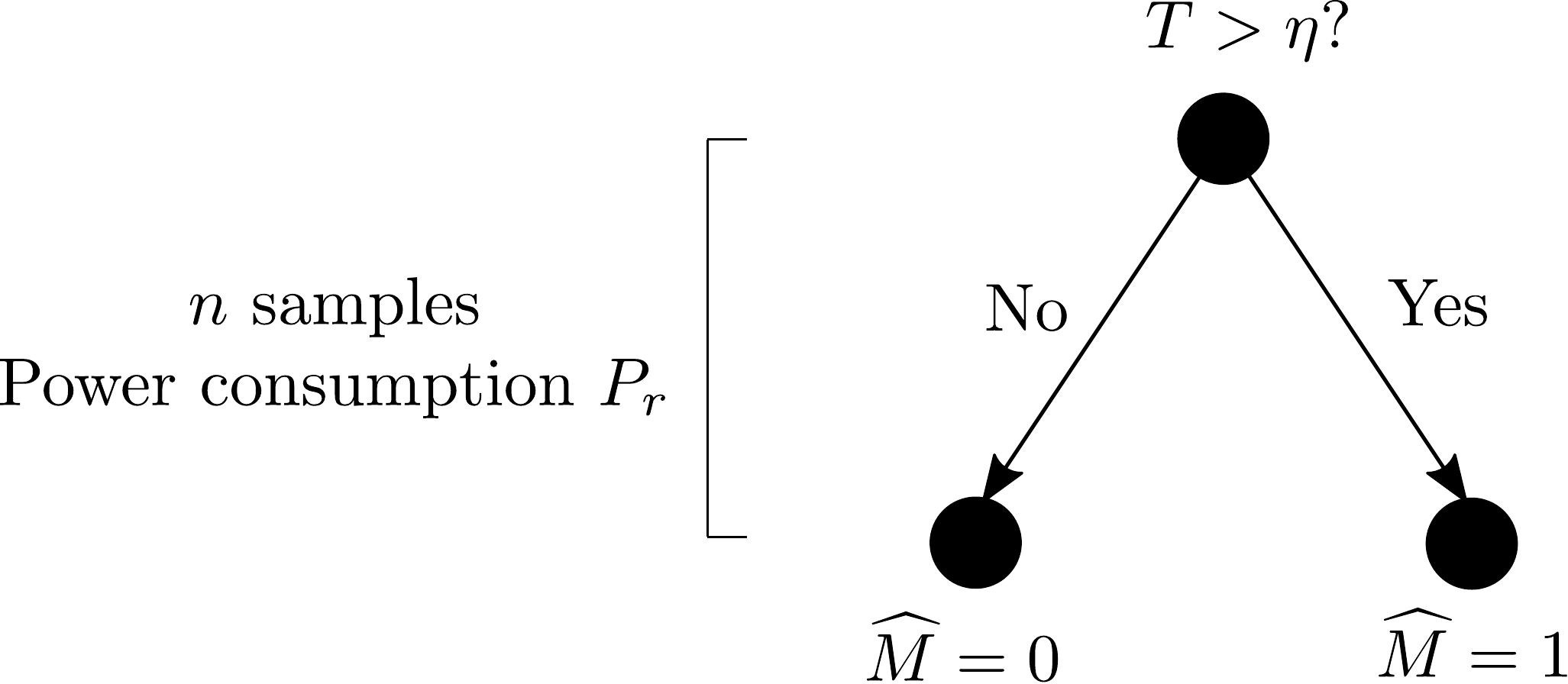}}
\caption{Single-phase scheme}
\label{fig:1p}
\end{figure}

The single-phase scheme is described in Fig.~\ref{fig:1p}. The receiver  performs an (optimal) LLR test on a fixed number of $n$ samples $Y_1,\ldots,Y_n$, each observed at a constant receiver power $P_r$. The test consists in declaring $\widehat{M}=1$ whenever $T\triangleq\frac{\sqrt{P}}{\sigma_r^2}\sum_{i=1}^nY_i-\frac{nP}{2\sigma_r^2}$ is above a given threshold $\eta$ that depends on the probability of false-alarm, and $\widehat{M}=0$ otherwise. 
The following theorem, which is essentially a direct consequence of Stein's Lemma \cite[Theorem~11.8.3]{cover} (see Appendix~\ref{app:asymp} for the proof),  states the performance of the single-phase scheme in the small probability of false-alarm regime: 
\begin{theorem}[Single-phase, asymptotics]\label{th:1p}
Suppose $\sigma_r^2=f(P_r)=kP_r^{-\gamma}$ for some fixed $\gamma\geq 1$ and $k>0$. For fixed $p_1,P,P_r>0$ and $0<\beta\leq 1$ the single-phase scheme achieves $P_{\text{Miss}}\leq\beta$ and
$$E= \frac{2 k }{P P_r^{\gamma-1}(1\pm o(1))}\ln \left(\frac{1}{P_{\text{FA}}}\right)$$
with a preamble length $n=O(\ln(1/P_{\text{FA}}))$, and  where $o(1)$ vanishes as $P_{\text{FA}}\to 0$.
\end{theorem}
By contrast with AdaSense, here $E$ does not decrease as communication gets sparser ({\it{i.e.}}, as $p_1$ decreases). 

\begin{figure}
\centering
\resizebox{0.49\textwidth}{!}{\includegraphics{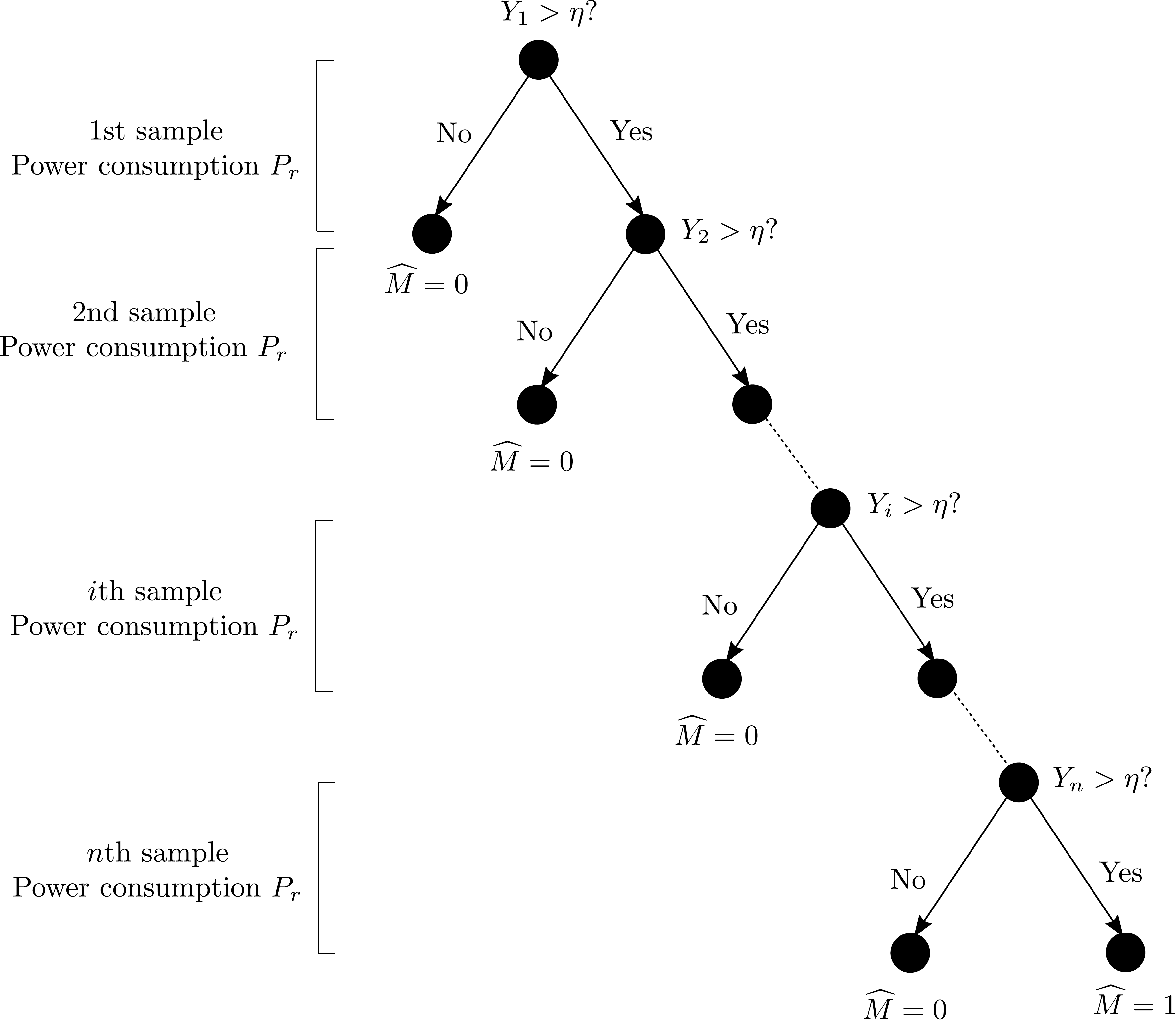}}
\caption{BMAC scheme}
\label{fig:bmac}
\end{figure}

An alternative scheme is the well-known BMAC scheme of \cite{BMAC} and described in Fig.~\ref{fig:bmac}. Similarly to the single-phase scheme, the receiver operates at a constant power. But instead of performing a hypothesis test based on all samples $Y_1,\ldots,Y_n$, the receiver performs a binary hypothesis test sequentially and independently on the $Y_i$'s and declares $\widehat{M}=0$ as soon as it finds a sample $Y_i\leq\eta$, where $\eta$ depends on the probability of false-alarm. If all $n$ samples exceed $\eta$, the receiver declares $\widehat{M}=1$.

The following theorem characterizes the performance of the BMAC scheme in the low probability of false-alarm regime (the proof is deferred to Appendix~\ref{app:asymp}):

\begin{theorem}[BMAC, asymptotics]\label{th:asympbmac}
Suppose $\sigma_r^2=f(P_r)={k}{P_r^{-\gamma}}$ for some fixed $\gamma\geq 1$ and $k>0$. For any fixed $p_1,P>0$, $n\geq 1$ and $0<\beta\leq 1$ the BMAC scheme achieves $P_{\text{Miss}}\leq \beta$ and
$$ E  \geq   (1-p_1)\left(\frac{2k}{nP}\ln\left(\frac{1}{P_{\text{FA}}}\right)\right)^{1/\gamma}(1-o(1))
$$
where $o(1)\to 0$ as  $P_{\text{FA}}\to 0$.
\end{theorem}

Similarly to the single-phase scheme, the energy spent by the BMAC scheme does not decrease as communication gets sparser. But note that the above lower bound has the preamble length $n$ in the denominator. It turns out that this is inherent to the BMAC scheme and is not due to a loose lower bound---see proof of Theorem~\ref{th:asympbmac}. A larger value of $n$ allows to lower the receiver power for a given level of false-alarm. However, if we impose BMAC to operate with the same preamble length as for AdaSense and BMAC and require $n=O(\ln(1/P_{\text{FA}}))$, AdaSense remains the most energy efficient in the sparse regime $p_1\to 0$, at a fixed but small probability of false-alarm. In the next section, we will also see that for a wide range of parameter values AdaSense performs the best in terms of energy consumption.

\begin{figure*}
\centering
\begin{subfigure}{0.45\textwidth}
  \centering
  \includegraphics[width=\linewidth]{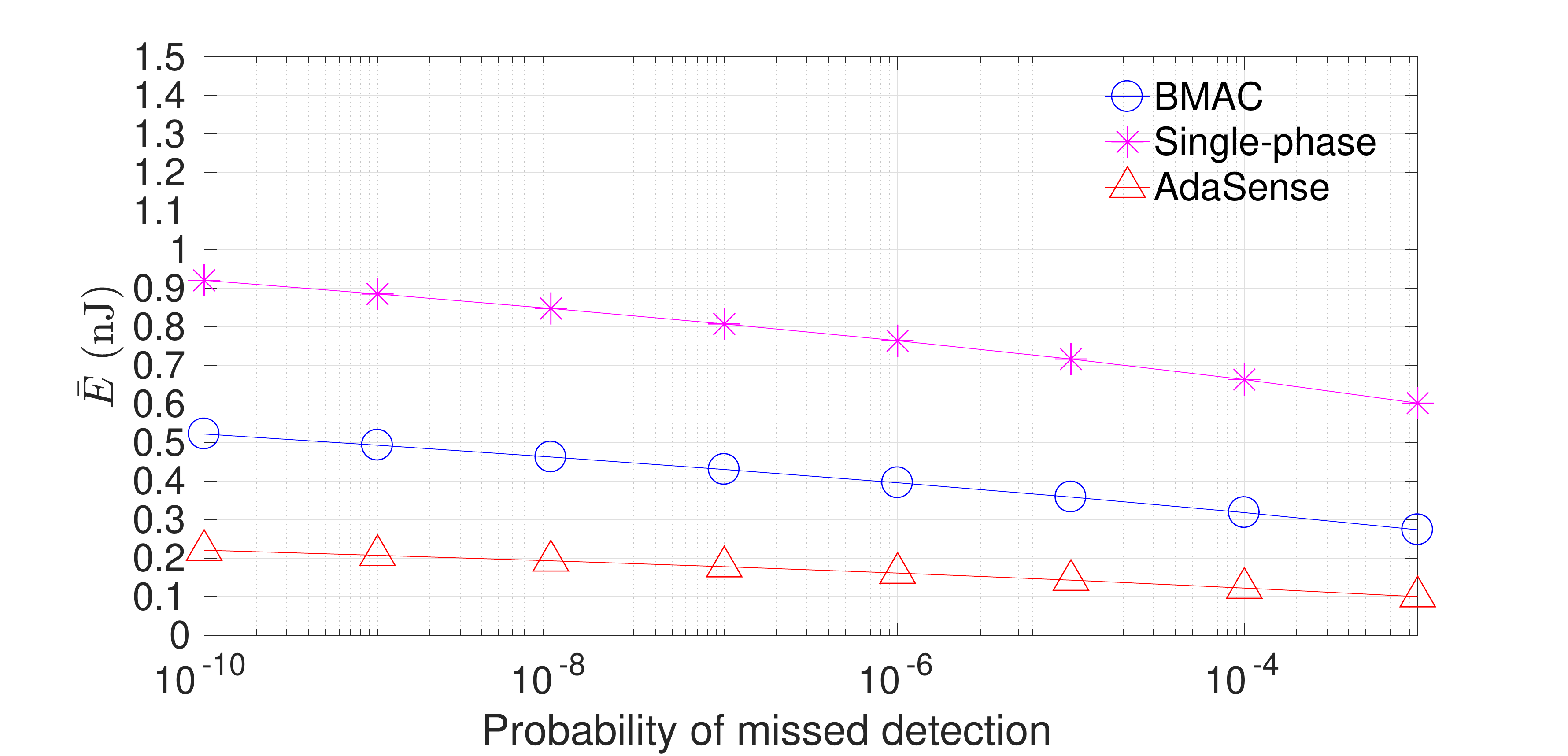}
  \caption{$n=30$, $P=-60$dBm, $P_{\text{FA}}=10^{-3}$.\\ Energy savings 60\%-63\%.}
  \label{fig:n30P60FA3}
\end{subfigure}
\begin{subfigure}{.45\textwidth}
  \centering
  \includegraphics[width=\linewidth]{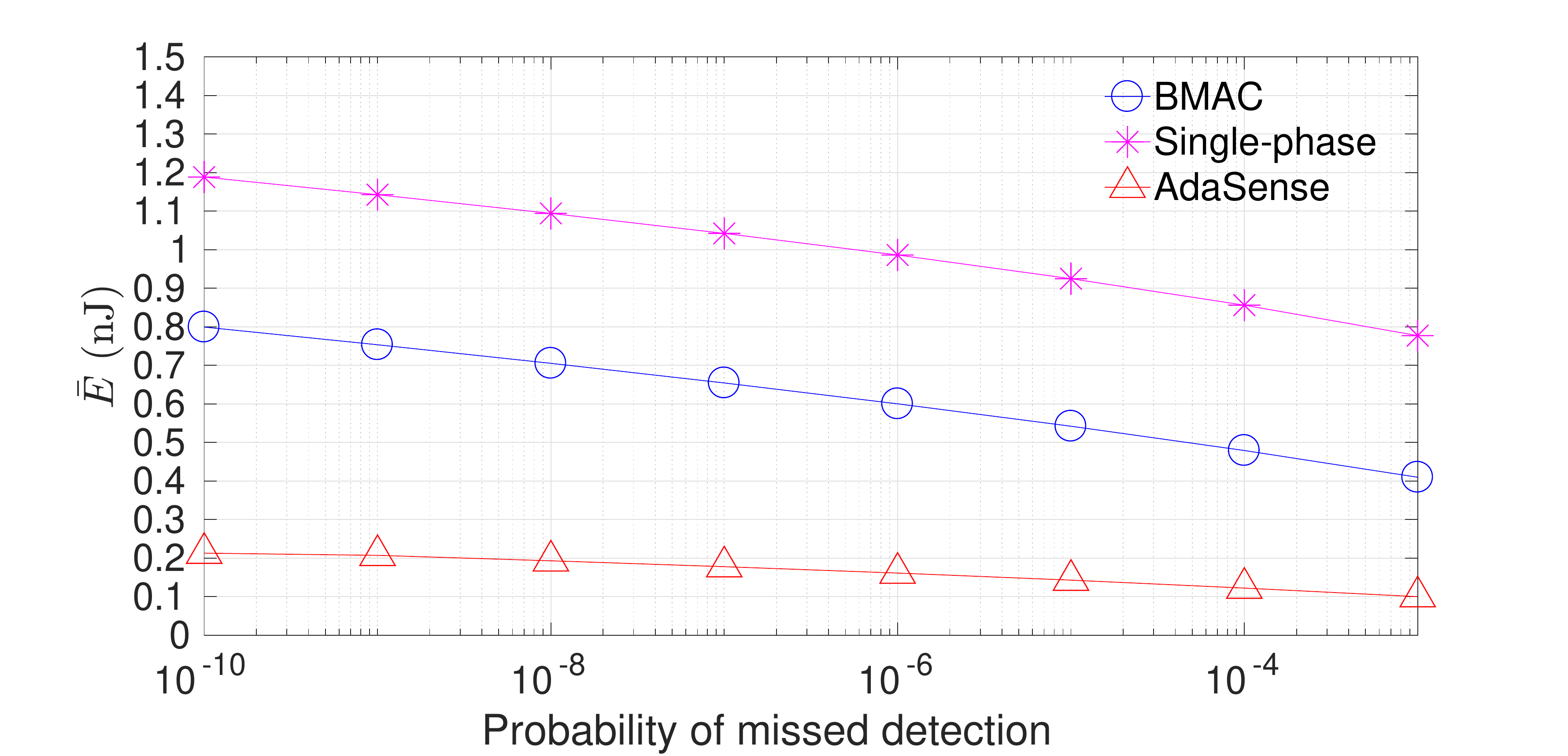}
  \caption{$n=50$, $P=-60$dBm, $P_{\text{FA}}=10^{-3}$.\\ Energy savings 73\%-76\%.}
  \label{fig:n50P60FA3}
\end{subfigure}
\medskip
\begin{subfigure}{.45\textwidth}
  \centering
  \includegraphics[width=\linewidth]{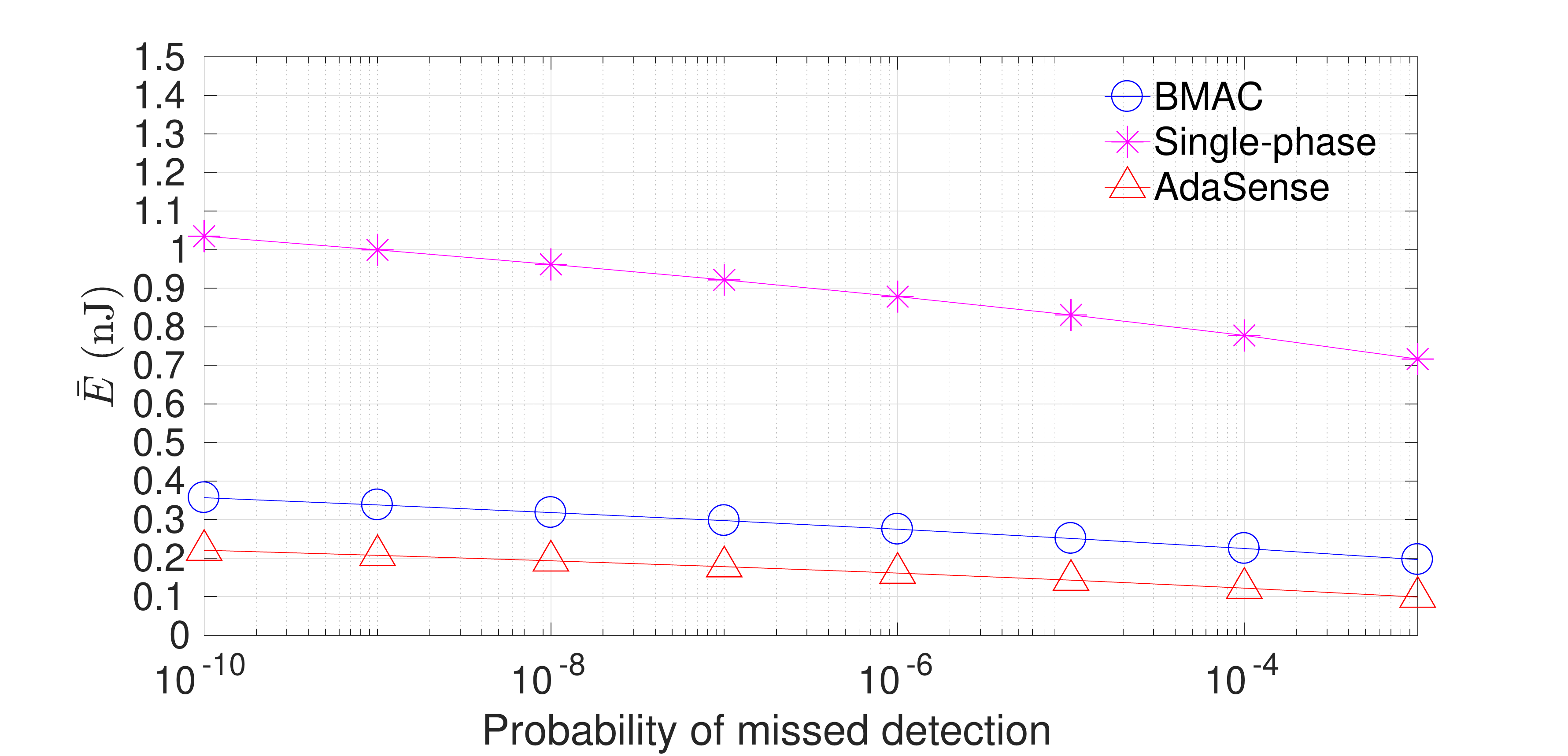}
  \caption{$n=30$, $P=-60$dBm, $P_{\text{FA}}=10^{-5}$.\\ Energy savings 40\%-50\%.}
  \label{fig:n30P60FA5}
\end{subfigure}
\begin{subfigure}{.45\textwidth}
  \centering
  \includegraphics[width=\linewidth]{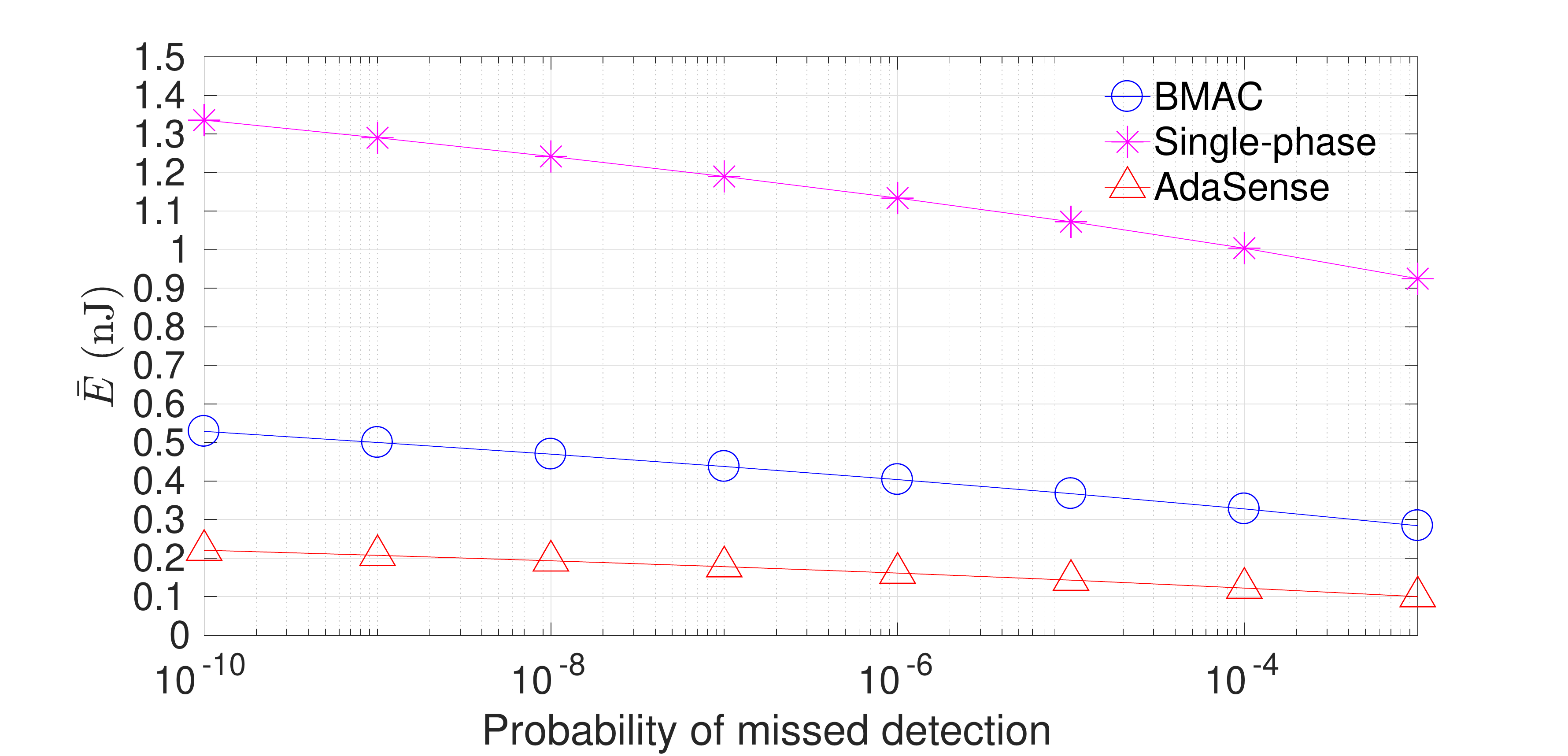}
  \caption{$n=50$, $P=-60$dBm, $P_{\text{FA}}=10^{-5}$.\\ Energy savings 59\%-65\%.}
  \label{fig:n50P60FA5}
\end{subfigure}
\medskip
\begin{subfigure}{.45\textwidth}
  \centering
  \includegraphics[width=\linewidth]{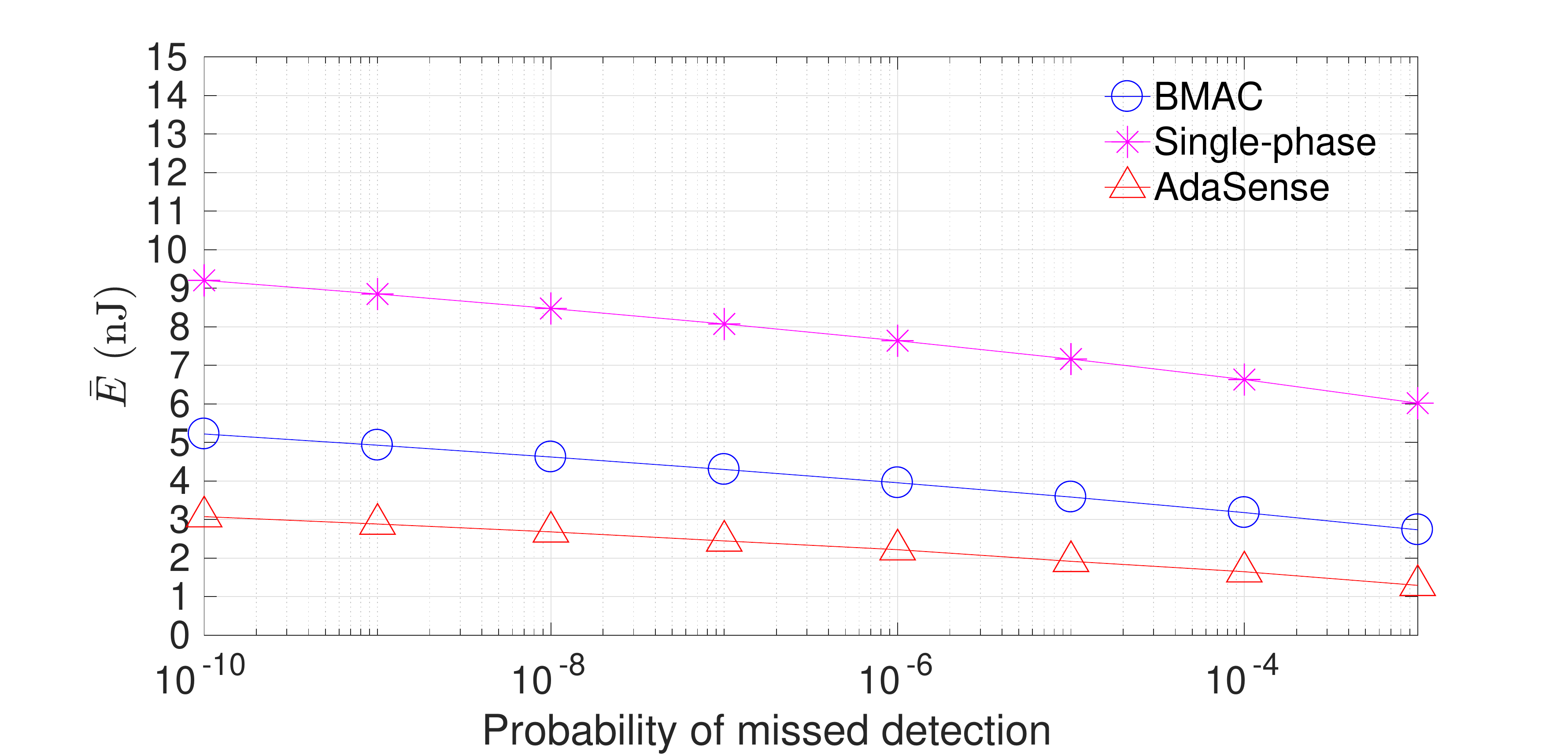}
  \caption{$n=30$, $P=-80$dBm, $P_{\text{FA}}=10^{-3}$.\\ Energy savings 42\%-52\%.}
  \label{fig:n30P80FA3}
\end{subfigure}
\begin{subfigure}{.45\textwidth}
  \centering
  \includegraphics[width=\linewidth]{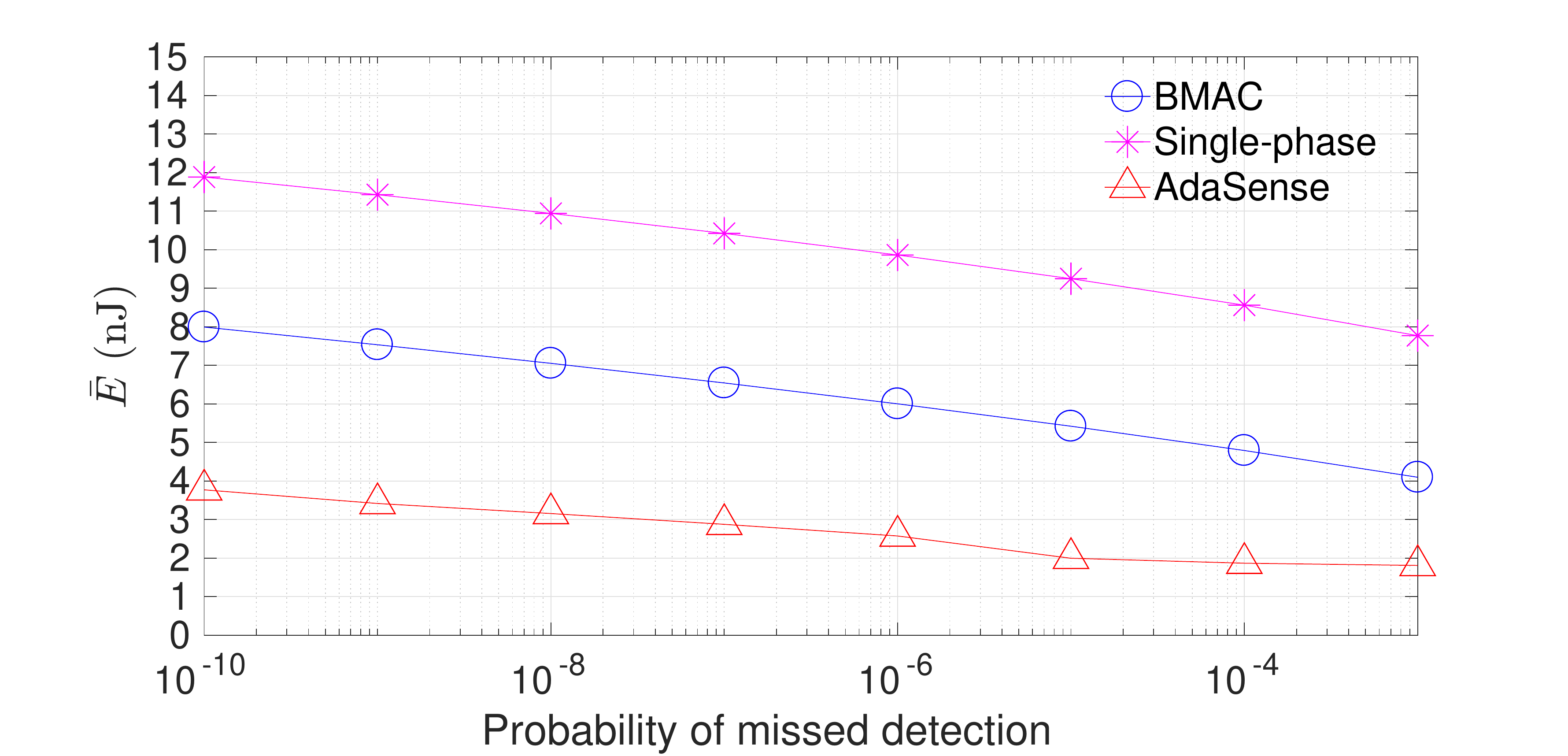}
  \caption{$n=50$, $P=-80$dBm, $P_{\text{FA}}=10^{-3}$.\\ Energy savings 53\%-62\%.}
  \label{fig:n50P80FA3}
\end{subfigure}
\medskip
\begin{subfigure}{.45\textwidth}
  \centering
  \includegraphics[width=\linewidth]{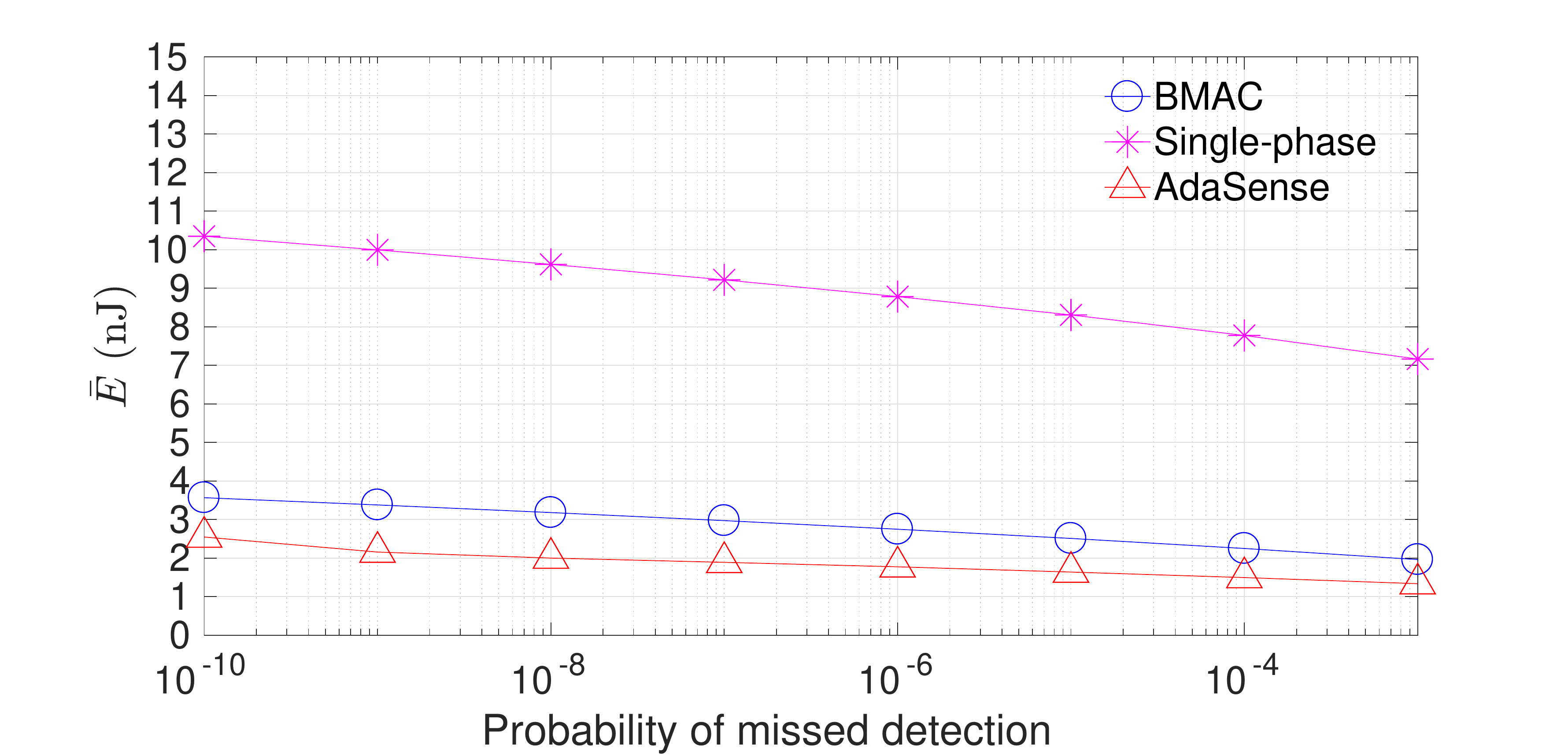}
  \caption{$n=30$, $P=-80$dBm, $P_{\text{FA}}=10^{-5}$.\\ Energy savings 29\%-32\%.}
  \label{fig:n30P80FA5}
\end{subfigure}
\begin{subfigure}{.45\textwidth}
  \centering
  \includegraphics[width=\linewidth]{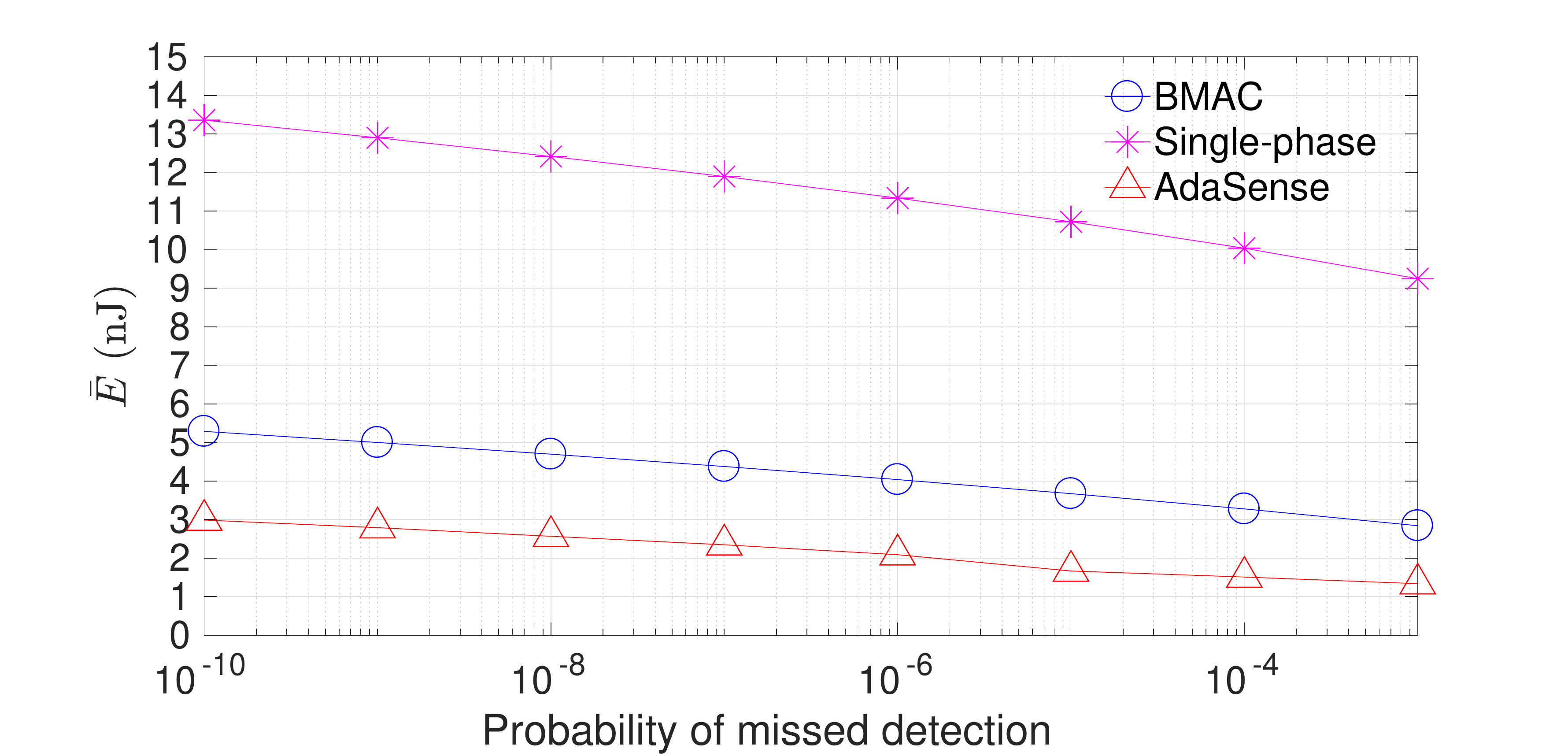}
  \caption{$n=50$, $P=-80$dBm, $P_{\text{FA}}=10^{-5}$.\\ Energy savings 44\%-53\%.}
  \label{fig:n50P80FA5}
\end{subfigure}
\caption{$E$ versus $P_{\text{Miss}}$ for the three different schemes with $p_1=10^{-10}$ and under different regimes of $P$, $n$, and $P_{\text{FA}}$. Energy savings compare AdaSense with the BMAC scheme---which is always more efficient than the single phase scheme for the considered parameters.}
\label{fig:comparison}
\end{figure*}

\subsection{\blue{Performance: Nonasymptotic Parameters regime}}\label{sec:compfin}

In this section we compare the performances of AdaSense, BMAC, and the single-phase scheme with parameters found in the context of wake-up receivers. Regarding preamble length we consider $n=30$ and $n=50$ which are consistent with the value in \cite{WUDC}. We then fix a target $P_{\text{FA}}$ equal to $10^{-3}$ or $10^{-5}$, as in \cite{Abe14,WUDC}, and plot $E$ as a function of $P_{\text{Miss}}$. 

\iffalse The theorems in the previous section show that given a targeted $P_{\text{FA}}$ and $P_{\text{Miss}}$, the average energy consumption $E$ depends on the maximum number of samples $n$ (or the preamble length), the probability of the presence of a message $p_1$, the received power per sample $P$, and the noise profile $f(\cdot)$, the values of which are noted and motivated as follows. \fi
 The probability  $p_1$ varies with applications. For example, temperature sensors deployed in some cities are monitored every few minutes \cite{city}, which, considering a standard bit rate of 100kbps and $n=50$, corresponds to $p_1=10^{-6}$. For other applications, for example fire alarms, $p_1$ can be much lower ({see \textit{e.g.}, \cite{BBJ07}}).

A $2017$ survey \cite{survey} recommends a worst-case received power of -83dBm, which corresponds to a line-of-sight communication over 1200m using an antenna transmitting at 10mW. Accordingly, we carry out the comparisons for a couple of values of received power $P=10^{-11}$ W, {\it{i.e.}}, $P=-80$ dBm and $P=10^{-9}$ W, {\it{i.e.}}, $P=-60$dBm. The noise profile, \textit{i.e.}, the function that relates the power consumption to the receiver noise variance, is unique to each receiver, and is determined by electrical simulations performed on the low noise amplifier to be used at the input stage of the receiver. For demonstration purposes, we use the noise profile determined in \cite[Section~V.B]{survey} by stating the tradeoff between \emph{sensitivity} and power consumption. Sensitivity is the defined as the received power needed to detect a single bit with a bit error rate of $10^{-3}$ \cite{NS14}. It is a textbook result that an SNR of 15dB is needed to detect a single bit with a bit error rate of $10^{-3}$ (see \cite[Example~II.B.2]{HVP}), and hence, receiver noise power is simply sensitivity lowered by 15dB. The sensitivity versus power consumption tradeoff mentioned in the survey can then be rephrased as follows: 1$\mu$W of power consumption leads to a receiver noise of $-55$dBm, and the noise power is decreased by 20dB by increasing the power consumption ten times. This tradeoff corresponds to the noise profile $f(\cdot)$ given by
$$
\sigma_r^2=f(P_r)=\frac{10^{-20.5}}{P_r^2},
$$
where $P_r$ is in Watts. The contribution of the thermal noise to the overall noise power is negligible, since the least value of $\sigma_r^2$ used in our comparisons is $-92.15$ dBm, whereas standard values of thermal noise power lie below $-113.83$ dBm.\footnote{The thermal noise power is calculated using the formula $\sigma_t^2=k_bTB$ in Watts \cite[Chapter~11]{Lee}, where $k_b$ is the Boltzman constant, $T$ is the temperature, and $B$ is the bandwidth. We have used $T=300$K (see for example \cite{NS14}) and $B=1$MHz, which is the maximum bandwidth allocated to low power devices in the IEEE 802.15.4 standard for low-rate communication \cite{802154}.}

In Fig.~\ref{fig:comparison}, we compare the exact energy consumption of the single-phase receive, the BMAC receiver, and AdaSense \eqref{eq:Ada:en}---after the optimization \eqref{eq:opt}---as a function of the probability of miss-detection. The exact performance of these schemes are deferred to Appendix~\ref{app:exact}, see Propositions~\ref{th:1pr}-\ref{th:exactada}. Each plot is for a specific set of parameters $n\in \{30,50\}$, $P\in \{-60 \text{ dBm}, -80\text{ dBm}\}$, and $P_{\text{FA}}\in \{10^{-3}, 10^{-5}\}$, and holds for any $p_1< 10^{-4}$---in this range of $p_1$ plots remain unchanged. For AdaSense, the numerical optimization of \eqref{eq:opt} was obtained through the MATLAB \emph{fmincon} optimizer run on an simple desktop (Intel{\textregistered} Core\textsuperscript{TM} i5-7200U CPU @ 2.50GHz × 4 with 16~GB of memory). The maximum wall-clock time needed to run the optimization was about $40$ seconds per set of target parameters.

As we see in Fig.~\ref{fig:comparison}, AdaSense is always the most energy-efficient scheme, followed by the BMAC scheme and the single-phase scheme. Savings vary from $30\%$ to $75\%$, depending on the regimes of $P$, $n$, $P_{\text{FA}}$, and $P_{\text{Miss}}$. We also note that higher values of $P$, $n$, $P_{\text{FA}}$, and $P_{\text{Miss}}$ result in higher savings, and can go up to 76\% (see Fig.~\ref{fig:n50P60FA3}). On the other hand, the least amount of energy savings we record is 29\% when $P=-80$dBm, $n=30$, $P_{\text{FA}}=10^{-5}$ and $P_{\text{Miss}}=10^{-10}$ (see Fig.~\ref{fig:n30P80FA5}). 

\section{Implementation}\label{sec:imp}
\begin{figure}
  \centering
  \includegraphics[width=0.7\linewidth]{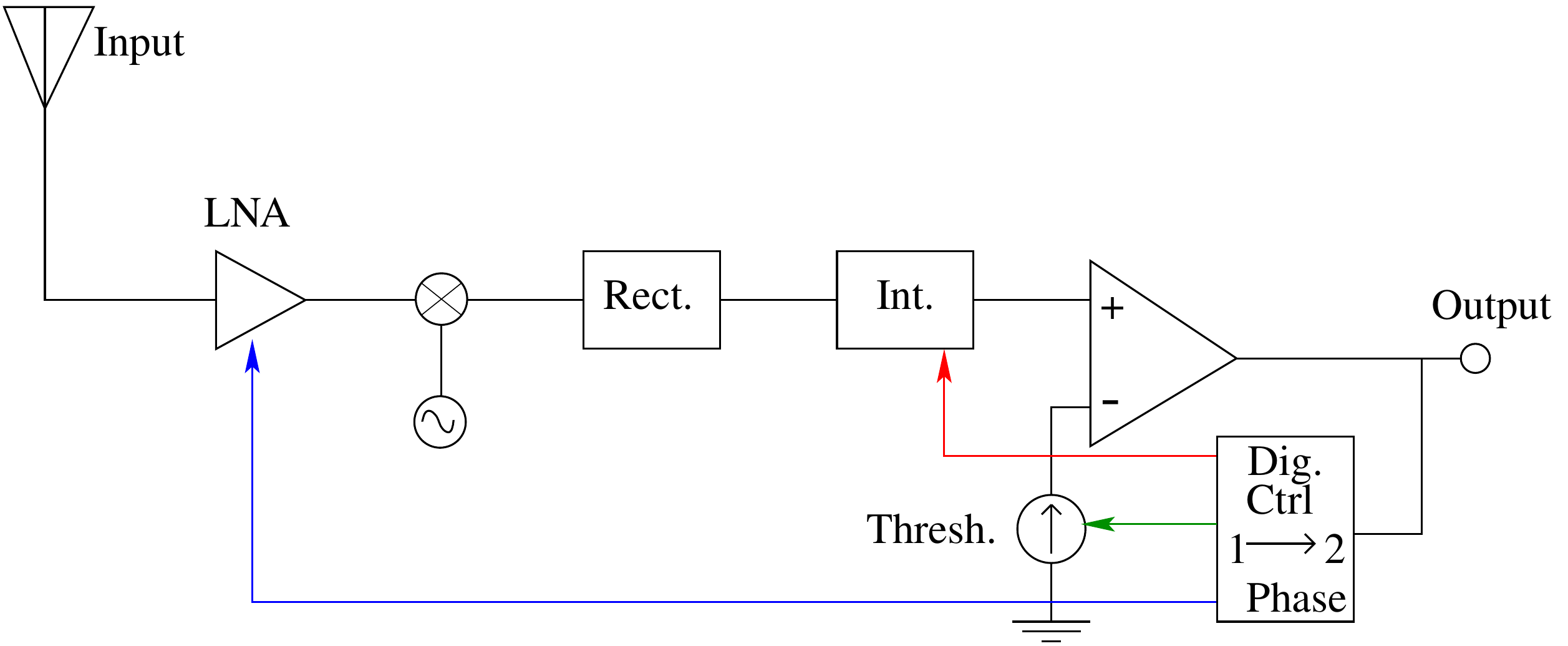}
  \caption{AdaSense implementation}
  \label{fig:arch}
\end{figure}

In this section we briefly outline a possible implementation of AdaSense, see Fig.~\ref{fig:arch}. The first few stages are standard for coherent receivers, which include a low noise amplifier (LNA) and  a down-mixer followed by a rectifier or envelope detector and the integrator, which convert the incoming BPSK modulated signal to the required weighted sum $\frac{\sqrt{P}}{\sigma_{r,1}^2}\sum_{i=1}^{\ell_1} Y_i$ (see Section~\ref{sec:ada}) \cite[Chapter~2]{VO}. This is then compared against the threshold $\eta_1+\frac{\ell_1P}{2\sigma_{r,1}^2}$ (see Section~\ref{sec:ada}) using a comparator. The heart of the architecture then lies in switching the mode of the receiver or turning it off, which is taken care of by the digital controller block. 

The overhead of AdaSense with respect to a classical non-adaptive receiver is the transition between the first and the second phase.  \blue{This requires modification of the noise figure (from $\sigma_{r,1}^2$ to $\sigma_{r,2}^2$), the integration constant which controls the weighted sum, and the threshold.
%The simplest approach to modify the noise factor is to change the bias current of the LNA amplifier, which reduces the noise added by the amplifier at the cost of a larger power consumption and vice versa. 
There are several efficient approaches to reconfigure the receiver noise figure. Firstly, note that the Friis formula (see \cite{Friis}) says that the components at the beginning of the receiver chain (\textit{i.e.}, the radio frequency (RF) components) are the major sources of the receiver noise. The low noise amplifier (LNA) occurring at the start of the receiver chain is therefore the most suitable site to perform the noise figure reconfiguration. The simplest approach to perform this reconfiguration is to adjust the LNA's bias current. A higher bias current leads to a higher transconductance of the LNA, and hence, a better noise figure \cite{Lee}, but at the cost of higher power consumption. Besides the bias current readjustment method, there exists other methods to reconfigure the noise figure of the LNA as well (see \cite{7598261, 6044942, 8341854}). The choice of the exact reconfiguration method will depend on the application and the receiver architecture.  It is worth mentioning that the narrow bandwidth of the modulated signals  relaxes the speed constraints on the reconfiguration and makes it easy to implement. To modify the integration constant we need to modify passive
elements (capacitors or resistors). This is achieved by implementing two different values of the considered elements (resistance or capacitance) and connecting or disconnecting them using switches depending on the phase.}  The comparison threshold is obtained through resistive bridge dividers. To generate different comparison voltages, it suffices to add an additional resistance in the resistive ladder, which allows us to generate two comparison voltages. Depending on the phase, the comparator is connected to one or the other using switches. 

%\textcolor{blue}{Chadi: Don't we need to reconfigure the integration constant too? }

\section{Concluding remarks}\label{sec:conc}
 In this paper we proposed a new channel sensing scheme which is particularly energy efficient in the sparse communication regime. The main difference with existing schemes is that AdaSense adaptively chooses the number of observed samples as well as the power at which these samples are observed. Notice that the results immediately extend to the case where the thermal noise is no longer negligible---just add the thermal noise variance $\sigma^2_t$ to the receiver noise $\sigma^2_r=f(P_r)$ throughout.  
 
 It should perhaps be emphasized that AdaSense operates only according to two power levels. In fact, one could envision a more gradual procedure where, after each observation the receiver either stops or observes the next sample at a potentially lower noise level. Preliminary calculations reveal that the energy gains provided by such a general scheme would not be substantial compared to AdaSense. Moreover, potential energy gains would probably be offset by an increased complexity at the hardware level. Indeed, as suggested in Section~\ref{sec:imp}, the implementation overhead of AdaSense with respect to a non-adaptive receiver appears to be negligible. This suggests that duty cycled receivers, including the duty-cycled \emph{wake-up receivers} \cite{WUDC,WUDC2,WUDC3,WUDC4}, could benefit from AdaSense.

\begin{appendices}

\section{Exact analysis}\label{app:exact}

According to the channel model \eqref{eq:channel}, the distribution of the samples $Y_i$ under either of the hypothesis $H_0$ or $H_1$ is i.i.d., and is given by
\begin{equation}
Y_i\sim
\begin{cases}
\mathcal{N}(0,\sigma^2), & \text{   given }H_0\\
\mathcal{N}(\sqrt{P},\sigma^2), & \text{   given }H_1.\label{eq:sdist}
\end{cases}
\end{equation} 
Therefore, the log-likelihood ratio (LLR) of $m$ i.i.d. samples $Y_1,Y_2,\ldots,Y_m$ is given by 
\begin{align}
T & = \log\biggl(\frac{\prod_{i=1}^m\frac{1}{\sqrt{2\pi\sigma^2}}\exp\bigl\{-\frac{1}{2\sigma^2}(Y_i-\sqrt{P})^2\bigr\}}{\prod_{i=1}^m\frac{1}{\sqrt{2\pi\sigma^2}}\exp\bigl\{-\frac{1}{2\sigma^2}(Y_i)^2\bigr\}}\biggr)\notag\\
                & = \frac{\sqrt{P}}{\sigma^2}\sum_{i=1}^mY_i-\frac{mP}{2\sigma^2}.\label{eq:1p:1}
\end{align}
Hence,
\begin{equation}
T\sim
\begin{cases}
\mathcal{N}\biggl(-\frac{mP}{2\sigma^2},\frac{mP}{\sigma^2}\biggr) & \text{ given }H_0\\
\mathcal{N}\biggl(\frac{mP}{2\sigma^2},\frac{mP}{\sigma^2}\biggr) & \text{ given }H_1.
\end{cases}\label{eq:1p:2}
\end{equation}

The next three propositions state the performance of the single-phase scheme, the BMAC scheme, and AdaSense, in the non-asymptotic regime.

Proposition~\ref{th:1pr} is  textbook material (see, {\it{e.g.}}, \cite[Example~II.D.1]{HVP}) and follows from \eqref{eq:1p:1} and \eqref{eq:1p:2}:
\begin{proposition}\label{th:1pr}
Given $n$, $P$, $\sigma_r^2=f(P_r)$, the single-phase scheme achieves
$$
P_{\text{FA}}= Q\biggl(\frac{\sigma_r\eta}{\sqrt{nP}}+\frac{1}{2}\sqrt{\frac{nP}{\sigma_r^2}}\biggr),
$$
$$
\text{and } P_{\text{Miss}}=Q\biggl(-\frac{\sigma_r\eta}{\sqrt{nP}}+\frac{1}{2}\sqrt{\frac{nP}{\sigma_r^2}}\biggr).
$$
 The energy consumed by the receiver is\footnote{Recall that the transmitted symbols have unit duration.}
\begin{align}
E=nP_r.\label{enfix}\end{align}
\end{proposition}

\begin{proposition}[BMAC]\label{th:exactbmac}
Given $p_1$, $n$, $P$, $\sigma_r^2=f(P_r)$, the BMAC scheme achieves
$$
P_{\text{FA}}=P_{e,0}^n, 
$$
$$
\text{and } P_{\text{Miss}}=1-(1-P_{e,1})^n,
$$
where $P_{e,0}=Q(\frac{\eta}{\sigma_r})$, $P_{e,1}=1-Q(\frac{\eta-\sqrt{P}}{\sigma_r})$, with $\eta\in{\mathbb{R}}$. The average energy consumed by the BMAC scheme is
\begin{align}
E & = P_r\biggl[\frac{p_1P_{\text{Miss}}}{1-(1-P_{\text{Miss}})^{1/n}}+\frac{(1-p_1)(1-P_{\text{FA}})}{1-P_{\text{FA}}^{1/n}}\biggr]\notag\\
& = P_r\frac{(1-P_{\text{FA}})}{1-P_{\text{FA}}^{1/n}}+o(1)\label{enbmac}
\end{align}
where $o(1)$ tends to zero as $p_1\to 0$.
\end{proposition}

Note that given $p_1$, $n$, and $P$, fixing any two values among $P_{\text{FA}}$, $P_{\text{Miss}}$, and $E$ specifies the third one.

\begin{IEEEproof}[Proof of Proposition~\ref{th:exactbmac}]
We have
\begin{align}
P_{\text{FA}} & = \text{Pr}(\widehat{M}=1|H_0)\notag\\
    & = \prod_{i=1}^n\text{Pr}(Y_i>\eta|H_0)\label{eq:bmac:1}\\
    & = Q\biggl(\frac{\eta}{\sigma_r}\biggr)^n,\label{eq:bmac:2}
\end{align}
where \eqref{eq:bmac:1} follows from the fact that the BMAC detection rule declares $\widehat{M}=1$ if and only if $Y_i>\eta$ for $i\in\{1,\ldots,n\}$ and where \eqref{eq:bmac:2} follows from \eqref{eq:sdist}. Similarly, we have
\begin{align}
P_{\text{Miss}} & = \text{Pr}(\widehat{M}=0|H_1)\notag\\
    & = 1-\prod_{i=1}^n\text{Pr}(Y_i>\eta|H_1)\label{eq:bmac:3}\\
    & = 1-Q\biggl(\frac{\eta-\sqrt{P}}{\sigma_r}\biggr)^n.\label{eq:bmac:4}    
\end{align}
\iffalse where \eqref{eq:bmac:3} follows from the BMAC detection rule and the fact that the $Y_i$s are i.i.d., and where \eqref{eq:bmac:4} follows from \eqref{eq:sdist}.
\fi 
For the average energy consumption $E$ we have
\begin{align}
E & = \sum_{i=1}^niP_r\text{Pr}(N=i)\notag\\
       & = \sum_{i=1}^niP_r\biggl[p_1\text{Pr}(N=i|H_1)+(1-p_1)\text{Pr}(N=i|H_0)\biggr].\label{eq:bmac:5}
\end{align}
We now calculate $\text{Pr}(N=i|H_0)$. For $i<n$, note that  $N=i$ if and only if $Y_j>\eta$ for all $j<i$, and $Y_i\leq\eta$. Thus, by \eqref{eq:sdist}, we have $$\text{Pr}(N=i|H_0)=Q\biggl(\frac{\eta}{\sigma_r}\biggr)^{i-1}\biggl(1-Q\biggl(\frac{\eta}{\sigma_r}\biggr)\biggr).$$ On the other hand, $N=n$ occurs if and only if $Y_j>\eta$ for all $j<n$. Therefore, $\text{Pr}(N=n|H_0)=Q\biggl(\frac{\eta}{\sigma_r}\biggr)^{n-1}$. Similarly, we have $$\text{Pr}(N=i|H_1)=Q\biggl(\frac{\eta-\sqrt{P}}{\sigma_r}\biggr)^{i-1}\biggl(1-Q\biggl(\frac{\eta-\sqrt{P}}{\sigma_r}\biggr)\biggr),$$ for all $i<n$, and $$\text{Pr}(N=n|H_1)=Q\biggl(\frac{\eta-\sqrt{P}}{\sigma_r}\biggr)^{n-1}.$$ Define $p\triangleq Q\biggl(\frac{\eta}{\sigma_r}\biggr)$ and $q\triangleq Q\biggl(\frac{\eta-\sqrt{P}}{\sigma_r}\biggr)$. Then, using \eqref{eq:bmac:5}
\begin{align}
E&=\sum_{i=1}^{n-1}iP_r\biggl[p_1q^{i-1}(1-q)+(1-p_1)p^{i-1}(1-p)\biggr]+nP_r\biggl[p_1q^{n-1}+(1-p_1)p^{n-1}\biggr]\notag\\
& = P_r\biggl[p_1\sum_{i=0}^{n-1}q^i+(1-p_1)\sum_{i=0}^{n-1}p^i\biggr]\notag\\
        & = P_r\biggl[p_1\frac{1-q^n}{1-q}+(1-p_1)\frac{1-p^n}{1-p}\biggr]\notag\\
        & = P_r\biggl[p_1\frac{P_{\text{Miss}}}{1-(1-P_{\text{Miss}})^{1/n}}+(1-p_1)\frac{1-P_{\text{FA}}}{1-P_{\text{FA}}^{1/n}}\biggr],\label{eq:bmac:8}
\end{align}
where \eqref{eq:bmac:8} follows from the definitions of $p$ and $q$ together with  \eqref{eq:bmac:2} and \eqref{eq:bmac:4}.
\end{IEEEproof}

\begin{proposition}[AdaSense]
\label{th:exactada}
Given $p_1$, $n$, $P$, $\sigma_r^2=f(P_r)$, the first and second phase lengths $\ell_1,\ell_2\geq 0$ such that {$\ell_1+\ell_2\leq n$}, the powers in the first and second phases $P_{r,1},P_{r,2}\geq 0$, and the thresholds of the hypothesis tests of the first and second phases $\eta_1,\eta_2\in \mathbb{R}$, AdaSense yields
$$
P_{\text{FA}}=Q\biggl(\frac{\sigma_{r,1}\eta_1}{\sqrt{\ell_1P}}+\frac{1}{2}\sqrt{\frac{\ell_1P}{\sigma_{r,1}^2}}\biggr)Q\biggl(\frac{\sigma_{r,2}\eta_2}{\sqrt{\ell_2P}}+\frac{1}{2}\sqrt{\frac{\ell_2P}{\sigma_{r,2}^2}}\biggr),
$$
and
\begin{equation*}
P_{\text{Miss}} = Q\biggl(-\frac{\sigma_{r,1}\eta_1}{\sqrt{\ell_1P}}+\frac{1}{2}\sqrt{\frac{\ell_1P}{\sigma_{r,1}^2}}\biggr)+\left(1-Q\biggl(-\frac{\sigma_{r,1}\eta_1}{\sqrt{\ell_1P}}+\frac{1}{2}\sqrt{\frac{\ell_1P}{\sigma_{r,1}^2}}\biggr)\right)Q\biggl(-\frac{\sigma_{r,2}\eta_2}{\sqrt{\ell_2P}}+\frac{1}{2}\sqrt{\frac{\ell_2P}{\sigma_{r,2}^2}}\biggr),
\end{equation*}
The average energy consumed is given by
\begin{align}\label{eno}
E=\ell_1P_{r,1}+p_c\ell_2P_{r,2},
\end{align}
where
\begin{align*}
{p_c} & =(1-p_1) Q\biggl(\frac{\sigma_{r,1}\eta_1}{\sqrt{\ell_1P}}+\frac{1}{2}\sqrt{\frac{\ell_1P}{\sigma_{r,1}^2}}\biggr)+p_1\biggl(1-Q\biggl(-\frac{\sigma_{r,1}\eta_1}{\sqrt{\ell_1P}}+\frac{1}{2}\sqrt{\frac{\ell_1P}{\sigma_{r,1}^2}}\biggr)\biggr)\\
    & {=Q\biggl(\frac{\sigma_{r,1}\eta_1}{\sqrt{\ell_1P}}+\frac{1}{2}\sqrt{\frac{\ell_1P}{\sigma_{r,1}^2}}\biggr)}+o(1),
\end{align*}
where $o(1)$ tends to zero as $p_1$ tends to zero. 

\end{proposition}

\begin{IEEEproof}[Proof of Proposition~\ref{th:exactada}]
We begin by evaluating the LLR for samples in the first and the second phases. Recall that the first batch of $\ell_1$ samples are received with noise variance $\sigma_{r,1}^2=f(P_{r,1})$, while the second batch of $\ell_2$ samples are received with noise variance $\sigma_{r,2}^2=f(P_{r,2})$. Let $T_1$ and $T_2$ denote the LLRs of the samples in the first and the second phases, respectively. By \eqref{eq:1p:1}, we have
\begin{align}
T_1 & = \frac{\sqrt{P}}{\sigma_{r,1}^2}\sum_{i=1}^{\ell_1}Y_i-\frac{\ell_1P}{2\sigma_{r,1}^2},\notag\\
T_2 & = \frac{\sqrt{P}}{\sigma_{r,2}^2}\sum_{i=\ell_1+1}^{\ell_1+\ell_2}Y_i-\frac{\ell_2P}{2\sigma_{r,2}^2}.\label{eq:2p:1}
\end{align}
From \eqref{eq:1p:2}
\begin{equation}
T_1\sim
\begin{cases}
\mathcal{N}\biggl(-\frac{\ell_1P}{2\sigma_{r,1}^2},\frac{\ell_1P}{\sigma_{r,1}^2}\biggr) & \text{ given }H_0\\
\mathcal{N}\biggl(\frac{\ell_1P}{2\sigma_{r,1}^2},\frac{\ell_1P}{\sigma_{r,1}^2}\biggr) & \text{ given }H_1,
\end{cases}\label{eq:2p:2}
\end{equation}
and
\begin{equation}
T_2\sim
\begin{cases}
\mathcal{N}\biggl(-\frac{\ell_2P}{2\sigma_{r,2}^2},\frac{\ell_2P}{\sigma_{r,2}^2}\biggr) & \text{ given }H_0\\
\mathcal{N}\biggl(\frac{\ell_2P}{2\sigma_{r,2}^2},\frac{\ell_2P}{\sigma_{r,2}^2}\biggr) & \text{ given }H_1.
\end{cases}\label{eq:2p:3}
\end{equation}
Note that $T_1$ and $T_2$ are independent since they are functions of different sets of samples.

We now proceed to evaluate $P_{\text{FA}}$ and $P_{\text{Miss}}$. Note that $\widehat{M}=1$, if and only if $T_1>\eta_1$ and $T_2>\eta_2$. Therefore,
\begin{align}
P_{\text{FA}} &= \text{Pr}(T_1>\eta_1|H_0)\text{Pr}(T_2>\eta_2|H_0)\label{defpa} \\
&= Q\biggl(\frac{\sigma_{r,1}\eta_1}{\sqrt{\ell_1P}}+\frac{1}{2}\sqrt{\frac{\ell_1P}{\sigma_{r,1}^2}}\biggr)Q\biggl(\frac{\sigma_{r,2}\eta_2}{\sqrt{\ell_2P}}+\frac{1}{2}\sqrt{\frac{\ell_2P}{\sigma_{r,2}^2}}\biggr),    \end{align}
where the second equality  follows from \eqref{eq:2p:2} and \eqref{eq:2p:3}. Similarly, we have
\begin{align*}
P_{\text{Miss}} &  =\text{Pr}(T_1\leq\eta_1|H_1)+\text{Pr}(T_1>\eta_1|H_1)\text{Pr}(T_2\leq\eta_2|H_1)\\
                &  = Q\biggl(-\frac{\sigma_{r,1}\eta_1}{\sqrt{\ell_1P}}+\frac{1}{2}\sqrt{\frac{\ell_1P}{\sigma_{r,1}^2}}\biggr)+\left(1-Q\biggl(-\frac{\sigma_{r,1}\eta_1}{\sqrt{\ell_1P}}+\frac{1}{2}\sqrt{\frac{\ell_1P}{\sigma_{r,1}^2}}\biggr)\right)Q\biggl(-\frac{\sigma_{r,2}\eta_2}{\sqrt{\ell_2P}}+\frac{1}{2}\sqrt{\frac{\ell_2P}{\sigma_{r,2}^2}}\biggr).
\end{align*}

Next, we compute the average energy consumption $E$. We have $$E=\ell_1P_{r,1}+p_c\ell_2P_{r,2}$$where $p_c$
denotes the probability that the scheme continues after the first phase by $p_c$, that is  $T_1>\eta_1$. From \eqref{eq:2p:2}, we have
\begin{align}
p_c  & = (1-p_1)\text{Pr}({T_1>\eta_1}|H_0)+p_1\text{Pr}({T_1>\eta_1}|H_1)\notag\\
& = (1-p_1) Q\biggl(\frac{\sigma_{r,1}\eta_1}{\sqrt{\ell_1P}}+\frac{1}{2}\sqrt{\frac{\ell_1P}{\sigma_{r,1}^2}}\biggr)+p_1\biggl(1-Q\biggl(-\frac{\sigma_{r,1}\eta_1}{\sqrt{\ell_1P}}+\frac{1}{2}\sqrt{\frac{\ell_1P}{\sigma_{r,1}^2}}\biggr)\biggr).\label{eq:2p:4}
\end{align} 
which concludes the proof.
\end{IEEEproof}

\section{Asymptotics}\label{app:asymp}

\begin{IEEEproof}[Proof of Theorem~\ref{th:1p} (Single-phase)] From Stein's Lemma {\cite[Theorem~11.8.3]{cover}}, if the single-phase satisfies $P_{\text{Miss}}\leq \beta$, for some fixed $0<\beta<1$, then\footnote{We write $f(n)=g(n)\pm q(n)$ if $g(n)-q(n)\leq f(n)\leq g(n)+q(n)$.} 
\begin{align}
P_{\text{FA}}
&= \exp\left({-\frac{P}{2}\frac{1}{\sigma^2_{r}} n\pm o(n)}\right)\nonumber \\
&= \exp\left({-\frac{P}{2k}P_{r}^\gamma n\pm o(n)}\right) \nonumber \\
&= \exp\left({-\frac{P P_{r}^{\gamma-1}}{2k}E\pm o(E)}\right)
\label{eq:asymp1p:pfa}
\end{align}
where for the first inequality we used the noise profile  $\sigma_r^2=kP_r^{-\gamma}$ and where the second inequality holds since $E=nP_r$ (see \eqref{enfix}). Therefore, 
$$E= \frac{2 k }{P P_r^{\gamma-1}(1\pm o(1))}\ln (1/P_{\text{FA}})$$
where $o(1)$ vanishes as $P_{\text{FA}}\to 0$---notice that as $P_{\text{FA}}\to 0$ we necessarily have $n$ and $E$ tend to infinity.  
\end{IEEEproof}
 Notice that the proof of Theorem~\ref{th:1p} also holds if $0<\gamma <1$. The same comment goes for the proof of Theorem~\ref{th:asympbmac} (see below). We nevertheless chose to state these theorems by restricting $\gamma\geq 1$ for the sake of uniformity with Theorem~\ref{th:asympada} and since the case $\gamma<1$ is less natural as previously alluded to.
\begin{IEEEproof}[Proof of Theorem~\ref{th:asympada} (AdaSense)]\label{app:asympada}
By Proposition~\ref{th:exactada}
 \begin{align}
    E=E_1+E_2 p_c
    \end{align}
where $$E_i=\ell_i P_{r,i}\quad i=1,2 $$  denotes the energy spent during the first and the second phase, and where $p_c$ denotes the probability that AdaSense performs the second phase, \textit{i.e.,} $p_c=\text{Pr}(T_1>\eta_1)$.

Now pick $0<\varepsilon<1 $ and  let us allocate the energy $E$ of the first and the second phase as 
$$E_1 =\varepsilon E=\ell_1 P_{r,1}$$
and \begin{align}\label{ener2}
    E_2 &=\frac{(1-\varepsilon)E}{ p_c}=\ell_2 P_{r,2}
    \end{align}
    Furthermore, pick \begin{align}\label{delta}\ell_i=\delta_i n \qquad i=1,2
\end{align}
such that $\delta_1>0$,  $\min\{E_2/n,\varepsilon\}\leq \delta_2 \leq 1-\varepsilon$, and $\delta_1+\delta_2\leq 1$.
As we will see below, the tradeoff between $P_{FA}$ and $E$ is improved as $\varepsilon$ gets smaller, that is when the overwhelming energy is spent during the second phase. 
    
 From Stein's Lemma \cite[Theorem~11.8.3]{cover} it follows that for any $\beta>0$, the thresholds $\eta_1,\eta_2$ may be set such that $P_{\text{Miss}}\leq \beta$ and such that \begin{align}\label{upb}\text{Pr}(T_i>\eta_i|H_0)\leq \exp\left({-\frac{P}{2}\frac{1}{\sigma_{r,i}^2}\ell_i+o(\ell_i)}\right) \quad i=1,2
 .\end{align}
 
  Hence, from \eqref{defpa}
\begin{align}
P_{\text{FA}}
&\leq \exp\left({-\frac{P}{2}\frac{1}{\sigma^2_{r,2}} \ell_2+o(\ell_2)}\right)\nonumber \\
&= \exp\left({-\frac{P}{2k}P_{r,2}^\gamma \delta_2 n+o(n)}\right)
\label{eq:asympada:pfa}
\end{align}
where for the second equality we used \eqref{delta} and the noise profile  $\sigma_r^2=kP_r^{-\gamma}$ . We now consider the problem of minimizing the first term in the exponent on the right-hand side of \eqref{eq:asympada:pfa} under the constraints \eqref{ener2}-\eqref{delta}, that is
\begin{align}\label{optim}
\max_{\substack{(P_{r,2},\delta_2):\\\min\{ E_2/n,\varepsilon\} \leq \delta_2 \leq 1-\varepsilon\\
 P_{r,2}\delta_2=E_2/n}}P_{r,2}^\gamma \delta_2
\end{align}
and distinguish the cases $\gamma=1$ and $\gamma>1$. 

\noindent {\emph{$\gamma=1$:}} In this case \eqref{optim} is equal to 
$(1-\varepsilon){E}/n p_c$ which together with \eqref{eq:asympada:pfa} yields 
\begin{align}
P_{\text{FA}}
\leq \exp\left({-\frac{(1-\varepsilon)P}{2k p_c}E(1+o(1))}\right)
\end{align}
where $o(1)$ tends to zero as $E$ tends to infinity.
Hence,
\begin{align}\label{erfatd}
    E\leq  p_c \frac{2k (1+o(1)) }{P(1-\varepsilon)}\ln\biggl(\frac{1}{P_{\text{FA}}}\biggr)
\end{align}
where $o(1)$ tends to zero as $P_{\text{FA}}$ tends to zero. 

\noindent {\emph{$\gamma>1$:}}
Here we obtain a tradeoff between energy and probability of false-alarm which is at least as good as for $\gamma=1$. To see this suppose $n\delta_2 P_{r,2} =E_2$ and pick $\delta_2>0$ small enough so that $ P_{r,2}\geq 1 $ (this is possible since by assumption $\delta_2$ can be taken as small as  $\min\{\varepsilon, E_2/n\}\leq E_2/n$). Then $P_{r,2}^\gamma \geq P_{r,2}$ and we deduce that \eqref{optim} is greater or equal to $E_2/n$. Hence \eqref{erfatd} also holds if $\gamma>1$.

\iffalse
\noindent {\emph{$\gamma<1$:}} By contrast with the case $\gamma>1$, the maximum value \eqref{optim} will be achieved for $\delta_2$ as large as possible, hence equal to $1-\varepsilon$. Then, from \eqref{ener2} and \eqref{delta} we get
$$E=np_cP_{r,2}$$
and \eqref{eq:asympada:2} becomes
\begin{align}P_{\text{FA}}\leq \exp\left(- \frac{P}{2k}\left(\frac{E}{n p_c}\right)^\gamma(1-\varepsilon+o(1))n\right)
\end{align}
where $o(1)$ tends to zero as $n$ (or $E$) tends to infinity. Hence,
\begin{align}\label{gammaless1}E\leq p_c\left(\frac{2k  n^{(\gamma-1)/\gamma}}{P(1-\varepsilon-o(1))}\ln\left(\frac{1}{P_{\text{FA}}}\right)\right)^{1/\gamma}
\end{align}
where $o(1)$ tends to zero as $P_{\text{FA}}$ tends to zero.
\fi
We now upper bound $p_c$ as 
\begin{align}\label{borp1}
     p_c&= \text{Pr}(T_1>\eta_1|H_0)(1-p_1)+\text{Pr}(T_1>\eta_1|H_1)p_1\nonumber \\
     &\leq \text{Pr}(T_1>\eta_1|H_0)+p_1 \nonumber\\
     &\leq p_1+o(1)
\end{align}
where $o(1)$ tends to zero as $n$ tends to infinity (or, equivalently, as $P_{\text{FA}}$ tends to zero), and where for the last inequality we used \eqref{upb} and the fact that $\ell_1$ increases with $n$. The theorem then follows from \eqref{erfatd} and \eqref{borp1}.
\end{IEEEproof}

\begin{IEEEproof}[Proof of Theorem~\ref{th:asympbmac}]\label{app:asympbmac}
From Proposition~\ref{th:exactbmac}, if BMAC satisfies $P_{\text{FA}}=\alpha$ and $P_{\text{Miss}}=\beta$, then
\begin{align}
\alpha & = Q\biggl(\frac{\eta}{\sigma_r}\biggr)^n \label{eq:bmac:alpha}\\
(1-\beta) & = Q\biggl(\frac{\eta-\sqrt{P}}{\sigma_r}\biggr)^n \label{eq:bmac:beta}\\
E & \geq  P_r \frac{(1-p_1)(1-\alpha)}{1-\alpha^{1/n}}. \label{eq:ebmac}
\end{align}
From \eqref{eq:bmac:alpha}, the noise profile $\sigma_r^2=kP_r^{-\gamma}$, and the Chernoff approximation on the Q-function $Q(x)=\exp(-x^2/2(1+o(1)))$ as $x\to \infty$ we deduce that
\begin{align}
P_r=\left(\frac{\sqrt{k} Q^{-1}(\alpha^{1/n})}{\eta}\right)^{2/\gamma}=\left(\frac{2k}{n\eta^2(1+o(1))}\ln(1/\alpha)\right)^{1/\gamma}
\end{align}
where $o(1)\to 0 $ as $\alpha\to 0$. Hence, from \eqref{eq:ebmac} 
\begin{align}
    E & \geq   \frac{(1-p_1)(1-\alpha)}{1-\alpha^{1/n}}\left(\frac{2k}{n\eta^2(1+o(1))}\ln(1/\alpha)\right)^{1/\gamma}
\end{align}
Assuming $\beta<1/2$ we  necessarily have $\eta\leq\sqrt{P}$ from \eqref{eq:bmac:beta} and we finally get\footnote{Notice that $\eta\leq\sqrt{P}$ is a crude bound and is definitely not sufficient to have $\beta<1/2$ as even for $\eta=\sqrt{P}$ we have $\beta=1-(1/2)^n\geq 1/2$.}
\begin{align}
    E & \geq   (1-p_1)\left(\frac{2k}{nP}\ln(1/\alpha)\right)^{1/\gamma}(1-o(1)).
\end{align}
as $\alpha\to 0$.
\end{IEEEproof}
\end{appendices}

\bibliographystyle{IEEEtran}
\bibliography{IEEEabrv,Draft}

%\begin{IEEEbiography}[{\includegraphics[width=1in,height=1.25in,clip,keepaspectratio]{manuj.png}}]{Manuj Mukherjee} (M'14) received the B.E. degree in Electronics and Telecommunication Engineering from 
%Jadavpur University, Kolkata, India, in 2011 and the M.S. (Engineering) and Ph.D. degrees in 
%Electrical Communication Engineering from the Indian Institute of Science, Bangalore, in 2017. Currently, he is a postdoctoral fellow at Telecom ParisTech, Paris, France. His research interests primarily lies in information theory and probability and their applications to real world scenarios. 
%\end{IEEEbiography}

%\EOD

\end{document}